\documentclass[useAMS,usenatbib]{mn2e}

\usepackage{graphicx}
\usepackage{epstopdf}
\usepackage{subfigure}
\usepackage{amsmath}
\usepackage{multicol}
\usepackage{multirow}
\usepackage{float}
\usepackage{color}

\makeatletter
\newcommand*{\rom}[1]{\expandafter\@slowromancap\romannumeral #1@}
\newcommand{\ha}{H$\alpha$}
\newcommand{\hb}{H$\beta+$[O\rom{3}]}
\newcommand{\oii}{[O\rom{2}]}
\newcommand{\Ld}{\mathcal{L}}
\providecommand{\e}[1]{\ensuremath{\times 10^{#1}}}
\makeatother
\title[\hb~ and \oii~ LFs out to $z \sim 5$]{Evolution of the \hb~and \oii~luminosity functions and the \oii~star-formation history of the Universe up to $z \sim 5$ from HiZELS}
\author[Khostovan et al.]{A.~A.~Khostovan$^{1}$\thanks{E-mail:
akhostov@gmail.com}, D.~Sobral$^{2,3,4}$, B.~Mobasher$^{1}$, P.~N.~Best$^{5}$, I.~Smail$^{6}$, J.~P.~Stott$^{7}$,  \newauthor S.~Hemmati$^{1}$, H.~Nayyeri$^{8}$\\
$^{1}$Department of Physics \& Astronomy, University of California, Riverside, United States of America\\
$^{2}$Instituto de Astrof\'{\i}sica e Ci\^{e}ncias do Espa\c{c}o, Universidade de Lisboa, OAL, Tapada da Ajuda, PT1349-018 Lisboa, Portugal\\
$^{3}$Departamento de F\'{i}sica, Faculdade de Ci\^{e}ncias, Universidade de Lisboa, Edif\'{i}cio C8, Campo Grande, PT1749-016 Lisbon, Portugal \\
$^{4}$Leiden Observatory, Leiden University, PO Box 9513, NL-2300 RA Leiden, the Netherlands\\
$^{5}$SUPA, Institute for Astronomy, Royal Observatory of Edinburgh, Blackford Hill, Edinburgh, EH9 3HJ, UK\\
$^{6}$Centre for Extragalactic Astrophysics, Durham University, South Road, Durham, DH1 3LE, UK\\
$^{7}$Institute of Computational Cosmology, Durham University, South Road, Durham, DH1 3LE, UK\\
$^{8}$Department of Physics \& Astronomy, University of California, Irvine, United States of America}
\begin{document}

\date{Accepted 2015 July 01.  Received 2015 May 29; in original form 2015 February 25}

\pagerange{\pageref{firstpage}--\pageref{lastpage}} \pubyear{2015}

\maketitle

\label{firstpage}

\begin{abstract}
We investigate the evolution of the \hb~and \oii~luminosity functions from $z \sim 0.8$ to $\sim5$ in four redshift slices per emission line using data from the High-{\it z} Emission Line Survey (HiZELS). This is the first time that the \hb~and \oii~luminosity functions have been studied at these redshifts in a self-consistent analysis. This is also the largest sample of \oii~and \hb~emitters (3475 and 3298 emitters, respectively) in this redshift range, with large co-moving volumes $\sim 1 \times 10^6$ Mpc$^{-3}$ in two independent volumes (COSMOS and UDS), greatly reducing the effects of cosmic variance. The emitters were selected by a combination of photometric redshift and color-color selections, as well as spectroscopic follow-up, including recent spectroscopic observations using DEIMOS and MOSFIRE on the Keck Telescopes and FMOS on Subaru. We find a strong increase in $L_\star$ and a decrease in $\phi_\star$ for both \hb~and \oii~emitters. We derive the \oii~star-formation history of the Universe since $z\sim5$ and find that the cosmic SFRD rises from $z \sim 5$ to $\sim 3$ and then drops towards $z \sim 0$. We also find that our star-formation history is able to reproduce the evolution of the stellar mass density up to $z\sim 5$ based only on a single tracer of star-formation. When comparing the \hb~SFRDs to the \oii~and \ha~SFRD measurements in the literature, we find that there is a remarkable agreement, suggesting that the \hb~ sample is dominated by star-forming galaxies at high-$z$ rather than AGNs.
\end{abstract}

\begin{keywords}
galaxies: evolution, galaxies: high-redshift, galaxies: luminosity function, mass function, cosmology: observations
\end{keywords}

\section{Introduction}
Our understanding of the mass assembly and star-formation processes of the Universe has improved greatly over the past few decades (for in-depth reviews, see \citealt{Kennicutt2012} and \citealt{Madau2014}). We currently have evidence to show that the cosmic star-formation rate (SFR) peaked at $z > 1$ and that about half of the current stellar mass density had been assembled by that time (e.g, \citealt{Lilly1996,Hopkins2006,Karim2011,Sobral2013}). However, many open questions remain. How fast did the star-formation rate density (SFRD) drop at $z > 2$? How has the population of star-forming galaxies changed over cosmic time? How does the evolution depend on the enviroment over cosmic time? To answer these questions, it is imperative that we use samples of star-forming galaxies that are low in contaminants and are well-defined in terms of selection methodology. 

There are many different star-formation indicators and calibrations in the literature. Each indicator traces the star-formation activity in galaxies independently and with different timescales. The ultraviolet (UV) light from bright, young stars with masses $> 5~M_\odot$ traces the bulk of the young population with time scales of $\sim 100$ Myr. UV light of stars with masses $> 10~M_\odot$ ionize the gas along the line-of-sight, resulting in the absorption and then re-emission of photons seen as nebular (e.g., Lyman, Balmer, and Paschen series of the Hydrogen Atom) and forbidden emission lines (e.g., [O\rom{3}] \& [O\rom{2}]). The lifetimes for stars capable of ionizing the surrounding gas and dust to form the nebular and forbidden emission lines are on the scale of $\sim 10$ Myr, allowing for the measurement of the instantaneous star-formation rate. Other indicators include far-infrared emission coming from the heating of dust shrouding the hot, UV bright, young stars and the synchrotron emission in the radio coming from accelerated electrons in supernovae. For an in-depth review of the various indicators and calibrations, we refer the reader to reviews in the literature (e.g., \citealt{Kennicutt1998}, \citealt{Calzetti2013}).

Despite the different indicators that exist, one can not say that only one of the indicators is the ``holy grail" of measuring the SFR of star-forming galaxies. However, using different tracers to map out the evolution of the cosmic SFR history is not the best solution either. This is because evolutionary studies based on samples selected in different ways and using different indicators/calibrations at different redshifts will be susceptible to complicated, strong biases and selection effects, which results in a significant scatter when combining all of them to probe the evolution of the cosmic SFR. Another issue is that most studies don't probe sufficiently large volumes to overcome the effects of cosmic variance. Furthermore, the effects of correcting for dust extinction, especially for UV and optical studies, can result in large systematic uncertainties. One requires an indicator that can be used to probe from the low-$z$ to the high-$z$ Universe using a robust and consistent methodology to reduce the effects and biases that come from making assumptions and differing selection techniques. 

Emission lines observed using narrow-band imaging techniques can provide an accurate and reliable sample of star-forming galaxies (e.g., \citealt{Bunker1995,Fujita2003,Glazebrook2004,Ly2007,Villar2008,Geach2008,Sobral2013}). The methodology utilizes two different images of the same field: one being from a broad-band filter and the other being from a corresponding narrow-band filter. The narrow-band image is dominated by emission-line galaxies and the continuum, while the broad-band image is dominated by the continuum with a small contribution from the emission-line. When the two are subtracted, the result is the removal of the continuum and an image of galaxies with emission-lines. The advantage of narrow-band imaging surveys is that they allow for the selection of emitters with a clean selection function by emission line flux and within a known narrow redshift range. This is because the filter width is quite narrow, such that any source brighter than expected from its broad-band magnitude is an emitter. Emission-line surveys via grism spectroscopy on the {\it HST} (e.g., \citealt{Colbert2013}) are also great accompaniments to narrow-band studies, as they are area-limited (area of the grism) while emission-line surveys are redshift-limited.

Most narrow-band surveys have focused on H$\alpha$ (e.g., \citealt{Tresse2002,Fujita2003,Pascual2005,Ly2007,Ly2011,Sobral2009,Sobral2013}) as it is a reliable star-formation indicator which is well-calibrated in the local universe and is only mildly affected by dust attenuation. The latest results of the High-Emission Line Survey (HiZELS, \citealt{Geach2008,Sobral2009,Sobral2012,Sobral2013}) have robustly traced the evolution of the cosmic SFR up to $z \sim 2$. This is the maximum redshift that H$\alpha$ surveys can probe from the ground, as at higher redshifts H$\alpha$ falls into the mid-IR and is blocked by water vapor and carbon dioxide in the atmosphere. To probe to higher-$z$ using the same narrow-band technique would require another emission line. The other major emission lines associated with star-forming galaxies are H$\beta$4861, [O\rom{3}]4959, [O\rom{3}]5007, and \oii3727 which can be probed up to $z \sim 3$ for \hb and up to $z \sim 5$ for \oii.

In the past decade, several H$\beta$, [O\rom{3}], and \oii~studies have been carried out (e.g., \citealt{Hammer1997,Hogg1998,Gallego2002,Hicks2002,Teplitz2003,Ly2007,Takahashi2007,Bayliss2011,Bayliss2012,Sobral2012,Ciardullo2013, Drake2013}), the majority of which had small sample sizes and, hence, suffered from cosmic variance biases. The majority observed up to $z \sim 1$, while only the works of \citet{Bayliss2011} and \citet{Bayliss2012} measured the \oii~SFR densities at $z \sim 2$ and $z \sim 4.6$, respectively. Both these works used small samples, with the $z \sim 4.6$ measurement having a sample size of only 3 \oii~emitters, greatly limiting any conclusion.

This paper presents, for the first time, the luminosity functions of \hb\footnotemark \footnotetext{Because the H$\beta$ and [O\rom{3}] emission lines are close to each other, photo-$z$ and color-color selections can not distinguish between them. The best way to fully differentiate the two is via spectroscopy. Based on line ratio studies, we can argue that most of the emitters will be [O\rom{3}] emitters, but to ensure we are not biasing our measurements based on such assumptions, we present the results as the combined measurement of \hb.} and \oii~emitters up to $z\sim5$ using the reliable selection techniques of \citet{Sobral2009,Sobral2012,Sobral2013} on the combined COSMOS and UDS narrow-band publicly available catalogs\footnotemark of HiZELS. The sample probes comoving volumes of up to $\sim 1 \times 10^6$ Mpc$^{3}$, which greatly reduces the effects of cosmic variance. This is also the largest sample of \hb~and \oii~emission line galaxies up to $z \sim 5$ to date in the literature and is used to effectively and robustly probe the evolution of the cosmic SFR density. 

\footnotetext{The narrow-band catalogs are available on VizieR and are from \citet{Sobral2013}.}

This paper is organized as follows: in Section 2, we outline the photometric and spectroscopic data sets that we use and the methodology utilized to effectively select \hb~ and \oii~emitters. In Section 3, we outline our volume calculations and the completeness and filter profile corrections. Also in Section 3 are the results of the luminosity functions. In Section 4, we present the results and discuss our cosmic star-formation rate densities and the evolution of the stellar mass density based on \oii~emitters. Section 5 outlines the conclusion of this work and is followed by appendix A, which presents our color-color selection criteria, and appendix B, which presents our binned luminosity function data points, and appendix C, which presents our SFR density compilation.

Throughout this paper, we assume $\Lambda$CDM cosmology with $H_0 = 70$ km s$^{-1}$ Mpc$^{-1}$, $\Omega_\mathrm{m} = 0.3$, and $\Omega_\Lambda = 0.7$ with all magnitudes presented as AB and the initial mass function is assumed to be a Salpeter IMF.

%% FILTER & REDSHIFT & VOLUME TABLE
\begin{table*}
\caption{A list of the narrow-band filters used \citep{Sobral2013}, along with the central wavelength (\micron) and the FWHM (\AA) of each filter. Included is the expected redshift of each emission line within the range of the FWHM of the filter and the comoving volume that is observed.}
\begin{tabular}{ l c c c c c c}
\hline
\hline
& & & \multicolumn{2}{c}{\hb5007} & \multicolumn{2}{c}{\oii3727}\\
\cline{4- 5}
\cline{6-7}
Filter & $\lambda_{\mathrm{obs}}$ & FWHM & $z$ & Volume & $z$ & Volume\\
 & (\micron) & (\AA) & & ($10^5$ Mpc$^{3}$ deg$^{-2}$) & & ($10^5$ Mpc$^{3}$ deg$^{-2}$)\\
 NB921 & 0.9196 & 132 & $0.84\pm0.01$ & 1.79& $1.47\pm0.02$& 3.75\\
 NBJ & 1.211 & 150 & $1.42\pm0.01$ & 3.11 & $2.25\pm0.02$ &4.83\\
 NBH & 1.617 & 211 & $2.23\pm0.02$ & 5.05 &$3.34\pm0.03$ & 6.53\\
 NBK & 2.121 & 210 & $3.24\pm0.02$ & 4.87 & $4.69\pm0.03$ & 5.68\\
\hline
\end{tabular}
\label{table:filter_probed}
\end{table*}

\section{Data}
\subsection{Selection Catalogs}
Our data consist of narrow-band and broad-band photometric data of the UDS \citep{Lawrence2007} and COSMOS \citep{Scoville2007} fields and spectroscopic follow-ups that are described in section \ref{sec:spectroscopy}. The catalogs that are described below are taken from \citet{Sobral2013} and are publicly available.

The narrow-band catalogs are from the High-{\it z} Emission Line Survey (HiZELS, \citealt{Geach2008,Sobral2009,Sobral2012,Sobral2013}). This project utilizes the narrow-band $J$, $H$, and $K$ filters of the Wide Field CAMera (WFCAM) on the United Kingdom Infrared Telescope (UKIRT) and the NB921 filter of the Suprime-Cam on the Subaru Telescope. Previous uses of this data focused primarily on H$\alpha$ emitting galaxies and their properties (e.g., \citealt{Garn2010,Geach2012,Sobral2010,Sobral2011,Sobral2013,Stott2013b,Stott2013a,Swinbank2012b,Swinbank2012a,Darvish2014}), but HiZELS is able to pickup more than just this emission line as it can detect any line above the flux limit. In this paper, we focus on the \hb~and \oii~emitting galaxies found by HiZELS. The broad-band catalogs are from the Cosmological Evolution Survey (COSMOS; \citealt{Scoville2007}) and the DR8 release of the Subaru-{\it XMM}-UKIDSS Ultra Deep Survey (UDS; \citealt{Lawrence2007}) catalog.\footnotemark \footnotetext{We refer the reader to the respective papers for further details on the creation of these catalogs. We also refer the reader to the UKIDSS and COSMOS websites for further information of the multi-wavelength photometric and spectroscopic data sets. (COSMOS: http://irsa.ipac.caltech.edu/data/COSMOS; UDS: http://www.ukidss.org)}

\subsection{Narrow-Band Selection of Potential Emitters}
In this subsection, we will review the methodology of selecting potential emitters followed by \citet{Sobral2013}. We refer the reader to this paper for a detailed overview. 

Potential emitters were selected by their color excess in terms of the parameter $\Sigma$ \citep{Bunker1995,Sobral2012}, which quantifies the significance of the excess of a source with respect to the random scatter expected for a source to have  a color excess $> 0$ in terms of their narrow-band magnitudes:
\begin{equation}
\Sigma = \frac{1 - 10^{-0.4(\mathrm{BB} - \mathrm{NB})}}{10^{-0.4 (\mathrm{ZP} - \mathrm{NB})} \sqrt{\pi r^2_\mathrm{ap} (\sigma^2_{\mathrm{NB}} - \sigma^2_{\mathrm{BB}})}}
\end{equation}
where BB and NB are the broad-band and narrow-band magnitudes, respectively, $r_{\mathrm{ap}}$ is the aperture radius in pixels, and $\sigma_{\mathrm{BB}}$ and $\sigma_{\mathrm{NB}}$ are the rms per pixel for the broad-band and narrow-band, respectively (see e.g. \citealt{Sobral2012,Sobral2013}). 

Emitters are selected on the basis that they have $\Sigma > 3$ and a rest-frame equivalent width EW$_0 > 25$ \AA. The second condition ensures that we select sources with a significant color excess for bright narrow-band magnitudes (see fig. 3 of \citealt{Sobral2013}). 

\begin{figure}
\centering
\includegraphics{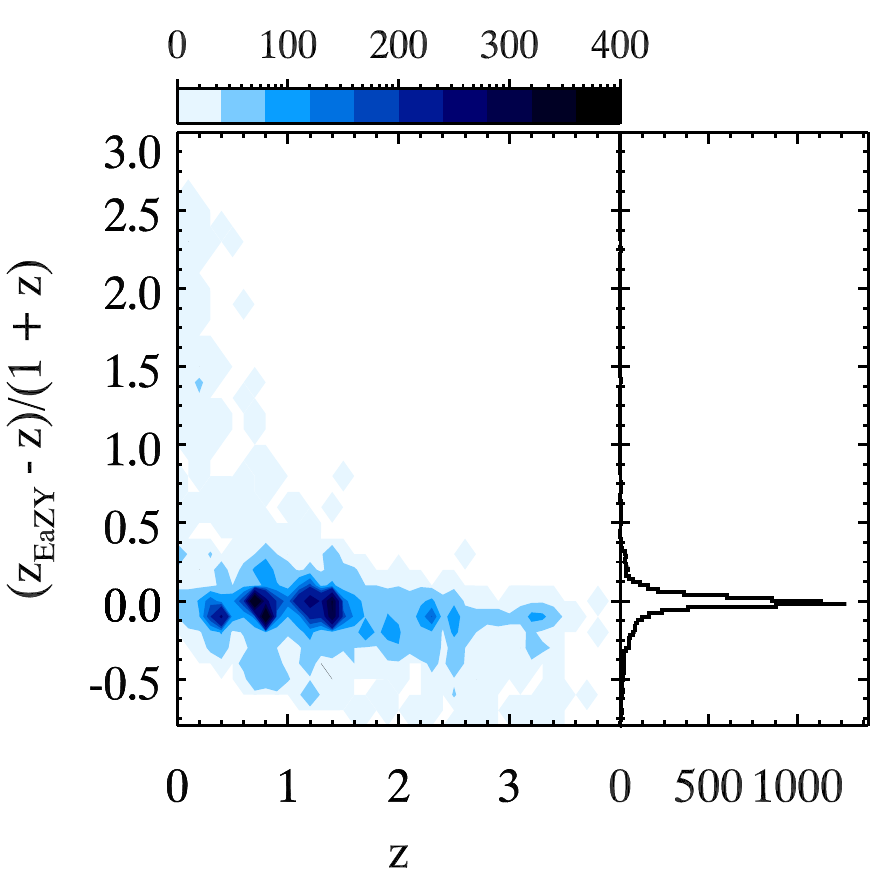}
\caption{The 2D density distribution of the errors between the redshifts determined by EaZY and those of \citet{Ilbert2009} (COSMOS, Le Phare) and \citet{Cirasuolo2007} (UDS, Hyperz) for all narrow-band excess sources (full catalog). We find that $\sim 95\%$ of our measurements are in agreement with the redshifts from the literature. The median error is measured as $0.038$ without sigma-clipping. There are over-densities at $z\sim0.4$, $0.8$, and $1.5$ that conform to redshifts for our major emission-lines. We find outliers with errors up to $\sim 2.5$, but these are only 4\% of our sample. In comparison to the wealth of spectroscopic data, we find a median photometric redshift error of $\Delta z/ 1 + z_{spec} = 0.047$ in comparison to spectroscopic redshifts.}
\label{fig:eazy}
\end{figure}

% Redshift Distribution Plot
\begin{figure*}
\centering
\includegraphics[scale=0.8,angle=90]{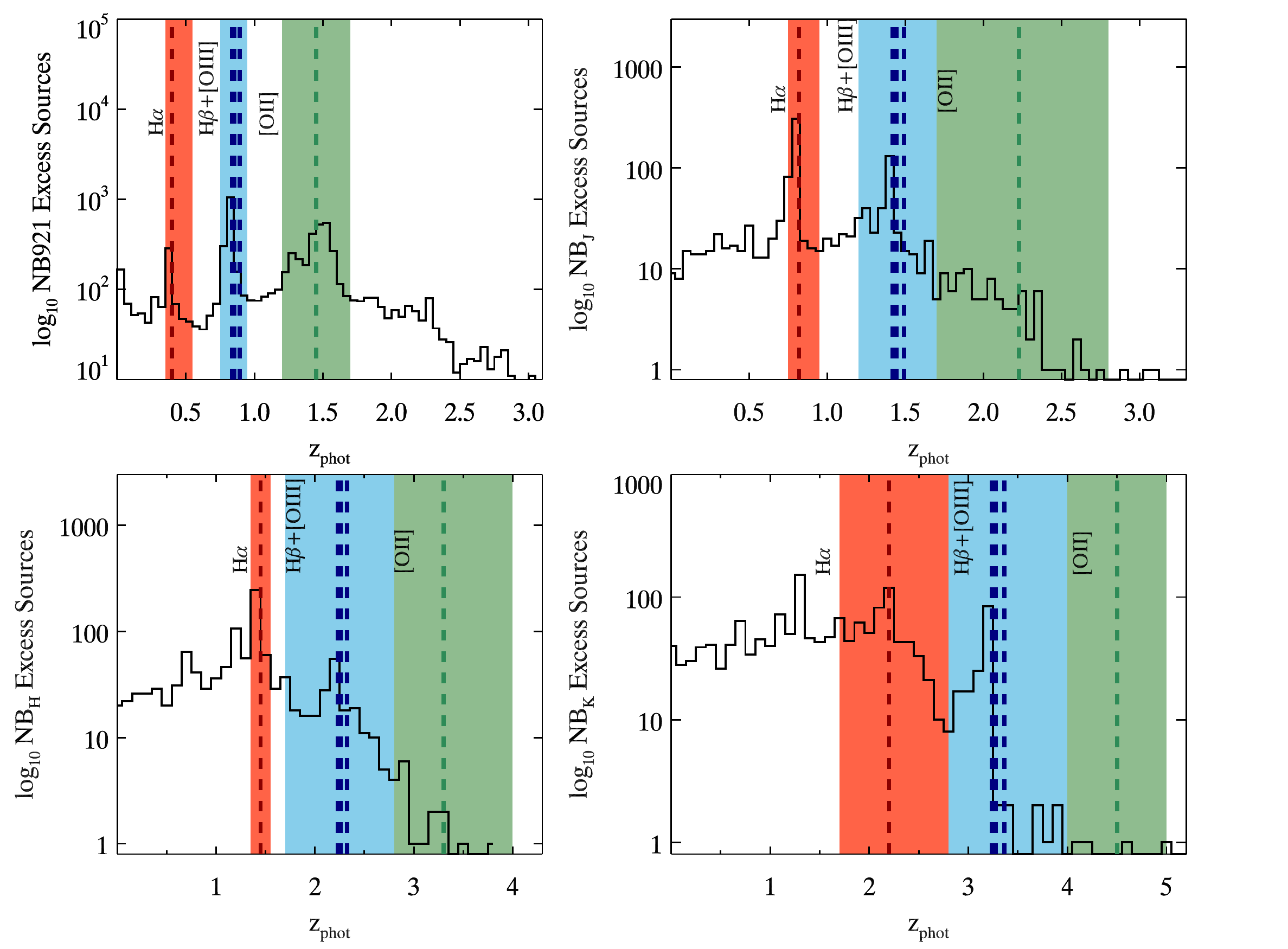}
\caption{Log-scale photometric redshift distributions of the emitters selected in NB921({\it upper left}), NB$_{\mathrm{J}}$({\it upper right}), NB$_{\mathrm{H}}$({\it lower left}), and NB$_{\mathrm{K}}$({\it lower right}). Each main peak is associated with a strong emission line, specifically H$\alpha$, \hb, and \oii. The {\it dashed red} line is the expected redshift of H$\alpha$ emitters. The {\it dashed blue} lines are the expected redshifts for \hb~emitters corresponding to H$\beta$4861, [O\rom{3}]4959, and [O\rom{3}]5007. It is clear then that differentiating these lines, even in narrow-band surveys, is quite difficult due to their close proximity to each other. Lastly, the {\it dashed green} line is the expected redshift of \oii~emitters. Highlighted in the corresponding, but lighter colors, are the photo-$z$ selection regions. The H$\alpha$ selection region is from \citet{Sobral2013}. The photo-$z$ distributions are from our photo-$z$ calculations using EaZY for the COSMOS and UDS fields.}
\label{fig:redshift_distributions}
\end{figure*}

\begin{table*}
\centering
\caption{ The number of emitters selected based on color-color, photo-$z$, spec-$z$, and dual/multi-emitters. We also include the total number of emitters found. It should be noted that an emitter can be selected by more than just one selection, such that by tallying up the number of emitters per column will result in a number larger than the number in the column that shows the total number of emitters. We also show the number of sources selected only based on color-color and only based on photo-$z$. This highlights the importance of having more than one selection technique. For example, if we solely relied on color-color selection, then we would have a loss of 21\% in our $z = 1.42$ \hb~sample. Furthermore, we also include the number of emitters selected as dual/multi-emitters. These are sources that were detected in more than one narrow-band, resulting in two or more detected emission-lines complementing each other (e.g., H$\alpha$ in NBH and \oii~in NB921 for $z \sim 1.47$).}
\begin{tabular}{c c c c c c c c c c}
\hline
\hline
Emission Line & Band & $z$ & Color-Color & CC Only & Photo-$z$ & Photo-$z$ Only & Spec-$z$ & Dual/Multi-Emitter & Total\\
\hline
{\bf \hb} & & & & & & & & &\\
 & NB921 & 0.84 & 2005 & 1000 & 1262 & 257 & 213 & 160 & 2477\\
 & NBJ & 1.42 & 277 & 41 & 314 & 78 & 15 & 23 & 371\\
 & NBH & 2.23 & 208 & 52 & 212 & 56 & 3 & 44 & 271\\
 & NBK & 3.24 & 145 & 11 & 158 & 24 & 2 & 0 & 179\\
\hline
{\bf \oii} & & & & & & & & & \\
 & NB921 & 1.47 & 3152 & 957 & 2211 & 16 & 97 & 213 & 3285\\
 & NBJ & 2.23 & 115 & 51 & 85 & 21 & 0 & 6 & 137\\
 & NBH & 3.30 & 29 & 18 & 16 & 5 & 1 & 0 & 35\\
 & NBK & 4.70 & 18 & 14 & 4 & 0 & 0 & 0 & 18\\
\hline
\end{tabular}
\label{table:total_emitters}
\end{table*}

\subsection{Photo-$z$ Measurements}
We initially used the photo-$z$ measurements from \citet{Ilbert2009} and \citet{Cirasuolo2007} for COSMOS and UDS, respectively, that were provided in the corresponding catalogs. The main problem of using those measurements is that in the UDS catalog, more than $>60\%$ of sources are without photometric redshifts. This raises issues when selecting emitters via redshift selection. Although the color-color selection is effective alone in selecting \hb~ and \oii~emitters (see below and appendices for discussion), we wish to have emitters selected by two independent methods to act as a check-and-balances to robustly select emitters. Therefore, we measured the photometric redshifts for all of the UDS and COSMOS narrow-band excess sources, using EaZY \citep{Brammer2008} to ensure that (1) the majority of the UDS catalog had reliable photometric redshifts, and (2) that both COSMOS and UDS had their redshifts determined by the same code, models, and assumptions. 

The filters we use for measuring photometric redshifts for our UDS sources are $UBVRizYJHK$ + {\it Spitzer} IRAC Ch1 - 4 + narrow-band (NB) + broad-band (BB) filters. For our COSMOS sources, we combine our $UBVgRizJK$+{\it GALEX} FUV \& NUV+{\it Spitzer} IRAC Ch1 - 4 +NB+BB catalog with that of \citet{Ilbert2009}. The benefit of this is that we include 12 Intermediate Subaru (e.g., IA427, IA464) bands and one Subaru NB711 band that were within the \citet{Ilbert2009} catalog. This results in a total of 29 filters to constrain the measurements. Furthermore, the benefit of using the narrow-band filters in measuring the redshifts is that these filters specifically capture emission-lines which are inherent in the SEDs. The Pegase13 spectral library that is used in EaZY includes a prescription for emission-lines, which makes the measurements more accurate when including the narrow-band filters. Figure \ref{fig:redshift_distributions} shows the benefit of using narrow-band filters in the fitting process by the sharp peaks for which we expect the major emission lines to be located in terms of redshift.

Figure \ref{fig:eazy} shows the 2D density distribution of $\sigma_z = (z_{\rm{EaZY}} - z)/(1+z)$, where $z$ are the redshifts measured by \citet{Ilbert2009} and \citet{Cirasuolo2007}, and the original photometric redshifts in the catalog. We find a median error of $0.038$ (all sources without sigma-clipping). Figure \ref{fig:eazy} also shows over-densities at $z \sim 0.4$, $0.8$, and $1.5$ which are expected as these are the most populated redshift slices in our sample, since they conform to major emission-lines. We find outliers up to $\sigma_z \sim 2.5$, but these only constitute a small fraction, such that $\sim 95\%$ of our sample are in agreement with \citet{Ilbert2009} and \citet{Cirasuolo2007}.

\subsection{Spectroscopic Redshifts}
\label{sec:spectroscopy}
We make use of the vast array of spectroscopic observations from the literature, which greatly enhances the reliability of our sample. In the COSMOS catalog, spectroscopic measurements are drawn from various studies as listed on the COSMOS website, as well as the $z$COSMOS measurements from \citet{Lilly2007}. The UDS catalog also includes measurements from various publications that are highlighted on the UKIDSS UKIRT website, including the UDSz survey \citep{Bradshaw2013, McLure2013}. We also include FMOS measurements from \citet{Stott2013b}, DEIMOS \& MOSFIRE measurements from Nayyeri et al., in prep, PRIMUS measurements from \citet{Coil2011}, and VIPERS measurements from \citet{Garilli2014}. In total, we have 1269 emitters that have spectroscopic redshifts with 661, 350, 177, and 81 emitters in NB921, NBJ, NBH, and NBK, respectively. This allows us to enhance the reliability of our sample and to test our photo-$z$ and color-color selections. In comparison to the spectroscopic redshifts, we have assessed the median errors of our photometric redshifts to be $\Delta z/1+z_{\rm{spec}} = 0.047$ (without sigma-clipping).

\subsection{Selection of \hb~and \oii~Emitters}
\label{sec:cc_selec}

The selection of potential \hb~and \oii~emitters is done by a combination of three different methods: (1) photometric redshift; (2) color-color; and (3) spectroscopic redshift. In this section, we will present the general selection criteria that we used to select our sample. For more detailed information about the specific selection cuts applied in each case, we refer the reader to appendix \ref{sec:selection_technique}. 

With three different selection methods, conflicts can arise where one selection method provides a result that conflicts with another method. To solve the issue, we prioritize the selection methods as such: 1) spectroscopic redshifts, 2) photometric redshifts, and 3) color-color. If the emitter has a spectroscopic redshift, then it is selected based only on measurement. If it doesn't have a spectroscopic redshift, then we select it based on its photometric redshift. Lastly, if the emitter has a photometric redshift but is not within the range of being selected as H$\alpha$ (see photo-$z$ selection criteria in \citealt{Sobral2013}), \hb, or \oii~or if the emitter does not have photo-$z$ measurements, then it is selected based on the color-color criterion. In most cases, we find that emitters with photo-$z$ within the redshift selection range are also found within the color-color selection area. In such cases, the emitters are selected based on both selection methods.

As shown in figure \ref{fig:redshift_distributions}, the local peaks in the photometric redshift distributions are located around the expected redshifts for \hb~and \oii~ emitters. This signifies that we have many \hb~and \oii~emitters in our sample. We select our emitter candidates by defining a range in the distribution of $z_{phot}$ that is centered on the expected redshift of the emission line, which is aligned with the peaks in figure \ref{fig:redshift_distributions}. In most cases, the number of H$\alpha$ emitters are the largest, followed by \hb~and \oii. In NB921 (figure \ref{fig:redshift_distributions}), \hb~and \oii~lines are the strongest, respectively, as these redshifts are near the peak of the cosmic star-formation history and also probe a much larger cosmic volume that H$\alpha$. Other populations of emission lines are found, such as Paschen series lines, He\rom{1}, and [S\rom{3}] (figure \ref{fig:redshift_distributions}).

Color-color selections are also applied for each redshift. The selection criteria used for $z \sim 1 - 3$ are from \citet{Sobral2013} as these are in perfect agreement with our large spectroscopic sample, while the criteria used for the \hb~and \oii~at $z = 3.3$ and \oii~ at $z = 4.5$ are based on the dropout color-color selection known as the Lyman break technique (\citealt{Dickinson1998,Stark2009}). For some redshifts, we use more than one color-color selection to reduce the contaminations from lower and higher-$z$ sources. All color-color selection definitions are found in table \ref{table:cc_sel}.

Whenever available, we select sources based on their spectroscopic redshifts. Sources for which the spectroscopic redshift contradicts the photo-$z$ and color-color selection are removed from the sample. By including spectroscopically confirmed sources, we increase the size of our sample. We also note that we use all our spectroscopic redshifts to confirm the robustness of our photo-$z$ and color-color selections.

Also in the selection process is selecting dual/multi-emitters. For some sources that are selected as emitters with some emission line, there also exists another emission line in another band. For example, an emitter that is selected in NBJ as an \hb~emitter ($z = 1.47$) can also be selected in NB921 as an \oii~emitter if observed in that narrow-band. \citet{Sobral2012} used this same technique in a double-blind study to find \oii~emitters in NB921 by using selected H$\alpha$ emitters in NBH as a proxy. The benefit to this technique is that it confirms emitters if they are selected in at least two bands; corresponding to getting a redshift out of two emission lines. In cases where we find dual/multi-emitters, we treat them as spectroscopic measurements as it is similar to having a spectroscopically confirmed emitter and include them in the sample.

One major source of contamination that we may have is having selected a source as an \oii~emitter when it is an \hb~emitter and vice versa. Also, there are situations where a source is selected as one of the emitters of interest, but also falls into the color-color selection for another emission line, or even H$\alpha$ (as these color-color selections were used by \citet{Sobral2009,Sobral2012,Sobral2013} to find such emitters). To overcome this degeneracy, we look at the photo-$z$ distributions shown in figure \ref{fig:redshift_distributions} as a probability distribution to assign the emission line based on the most probable line in the sample. In all cases, except for NB921, H$\alpha$ is the most probable line. For NB921, \hb~is the most probable.

Another source of contamination is from misidentified lines. From the wealth of spectra that we have, the majority of misidentified lines are H$\alpha$. Further details on the types of misidentified lines can be found in appendix \ref{sec:selection_technique}. Our total sample size is outlined in table \ref{table:total_emitters}.

%%%%%%%%%%%%%%%%%%%%%%%%%%%%%
% LUMINOSITY FUNCTION AND EVOLUTION SECTION%
%%%%%%%%%%%%%%%%%%%%%%%%%%%%%
\section{Luminosity Functions and Evolution}
We use the traditional $V_{max}$ estimator to create our binned data. The binned data is defined such that:
\begin{align}
\phi (L_j) = \frac{1}{\Delta \log L_{b,j}} \sum_{\log L_{c,j} - \Delta \log L_{b,j} /2.}^N \frac{1}{C(L_i) V_{max,i}}
\end{align}
where $L_j$ is the $j^{th}$ luminosity bin, $\Delta \log L_{b,j}$ is the bin-size, $\log L_{c,j}$ is the central luminosity of the $j^{th}$ bin, $C(L_i)$ being the completeness of the $i^{th}$ source described in section \ref{sec:completeness}, and $V_{max,i}$ being the volume for which that source may be detected as described in section \ref{sec:volume}.

\begin{figure}
\centering
\includegraphics[scale=0.75]{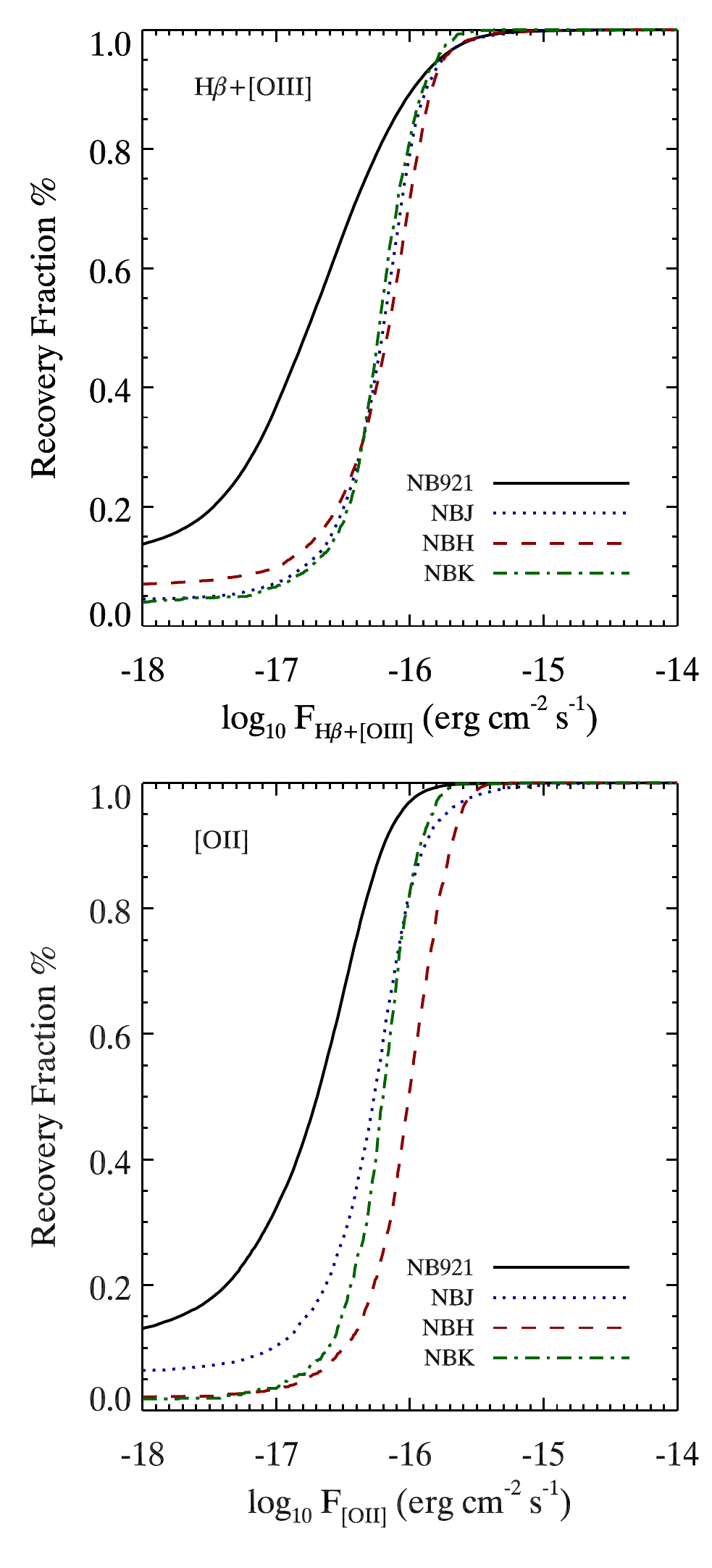}
\caption{Average line completeness for the entire sample per band. Note that the completeness does vary between the COSMOS and UDS fields, as well as from image-to-image within each field. This is because each image has a different depth. We compute the completeness based on each image to account for this discrepancy.}
\label{fig:completeness}
\end{figure}

\subsection{Line Completeness}
\label{sec:completeness}

To assess the completeness, we follow the methodology proposed in \citet{Sobral2012,Sobral2013}. We start with the full catalog that has all the emitters and includes sources which did not make our emitter selection. To measure the recovery fraction based on the emission line flux, we input a mock line flux starting at $10^{-18}$ erg s$^{-1}$ cm$^{-2}$ to all sources in the catalog. We then apply the excess selection criteria ($\Sigma$ and EW cuts) used in \citet{Sobral2013} followed by our color-color selections. The recovery fraction is then defined as the number of sources recovered divided by the total number of sources in the catalog. This is then repeated after increasing the input mock line flux by small increments. The expected result is that at $10^{-18}$ erg s$^{-1}$ cm$^{-2}$ the recovered fraction will be low and will increase as the input line flux increases. Figure \ref{fig:completeness} shows the average completeness determined for our \hb~and \oii~sources in the different narrow bands. The advantage of this technique is that we are not limiting our determination of the completeness to a certain model, but actually using the observed data itself to get the completeness correction. 

There are some important points to be noted. First, these simulations are run separately per image. This is because the depths of each image are not the same and thus can not be used together all at once to determine the completeness correction. Secondly, we apply an uncertainty of 20\% of the completeness correction to the other uncertainties in quadrature in order to take into account the errors associated with this method of determining the corrections.

\subsection{Volumes \& Filter Profile Correction}
\label{sec:volume}
We calculate the volumes assuming a top-hat filter that has the same range as the FWHM of the actual filter. We report the probed comoving volume per square degree in table \ref{table:filter_probed}. The volumes for each log-luminosity bins in the luminosity functions are reported in tables \ref{table:hbeta_lf_observed} and \ref{table:oii_lf_observed} for \hb~and \oii, respectively.

Although a top-hat filter makes our calculations easy, it is not a true representation of the throughput of the filter, which requires us to apply a filter profile correction. There are two main effects that the filter profile correction takes into account. The first is the {\it flux loss due to emitters that are close to the edge of the filter's FWHM}. Bright emitters at the wings will have a significant flux loss (close to 40\%; depends on the filter) and would be considered as a faint source. This gives an overall bias in our sample population of faint sources and a lack of bright sources. The second effect is {\it the volumes are corrected for the bright sources}. Any faint source that is close to the wings of the filter will most likely not be in our sample, but bright sources will be detected as faint emitters. This then implies that our bright emitters cover a wider range of the filter, meaning a wider range in redshift, and, therefore, a larger volume. 

We use the method proposed in \citet{Sobral2009, Sobral2012, Sobral2013}. We correct for the filter by creating a mock sample of $10^5$ fake emitters based on the luminosity function with the assumption of a top-hat filter. Random redshifts are assigned to each source and in a range covering the full filter profile, but not large enough that evolutionary effects of the luminosity function and cosmological structure biases the results. We assume a uniform redshift distribution. This mock sample will have the same distribution as the input top-hat luminosity function which we define as $\phi_{\mathrm{TH}}$. We then make a second mock sample with the same top-hat luminosity function but now apply the actual filter. This is done by:
\begin{align}
L_{corr} = \frac{\int_{z_1}^{z_2} L_{in}(z) T(z) dz}{\int_{z_1}^{z_2} T(z)dz}
\end{align}
where $T(z)$ is the filter-response function in terms of redshift and $L_{in}(z)$ is the luminosity of the emitter which is defined as $L_{in}(z) = L ~\delta(z_{\mathrm{rand}}- z_{\mathrm{filter}})$, where $z_{rand}$ is the randomly assigned redshift and $z_{\mathrm{filter}}$ is the matching redshift of the filter. This results in the loss of sources as some sources will have a redshift outside the range of the filter's FWHM. The luminosity function from this population is defined as $\phi_{\mathrm{Filter}}$ and is compared to $\phi_{\mathrm{TH}}$ in order to get the filter profile correction factor. The result shows that the bright sources are underestimated as expected, which changes the shape of the final luminosity function slightly. Specifically, it decreases the faint-end slope and increases the $L_\star$ and $\phi_\star$ slightly. This correction is applied to the LFs by dividing all the binned $\Phi(L)$ data points by the correction factor.

\subsection{Luminosity Function Fitting}
There are several different functions that have been proposed in the literature to describe the observed luminosity function of the Universe (see \citealt{Johnston2011} for an in-depth review). We adopt the most-widely accepted Schechter function to fit the observed luminosity function. The Schechter function is defined in its log-form as:
\begin{equation}
\Phi(L) \mathrm{d} L = \phi_\star ~\ln 10 ~\Bigg (\frac{L}{L_\star}\Bigg)^{1+\alpha} e^{-(L/L_\star)}\mathrm{d} \log_{10} L
\end{equation}
where $\phi_\star$ is the normalization of the luminosity function, $L_\star$ is the characteristic luminosity, and $\alpha$ is the faint-end slope. 

\begin{figure*}
\centering
\includegraphics{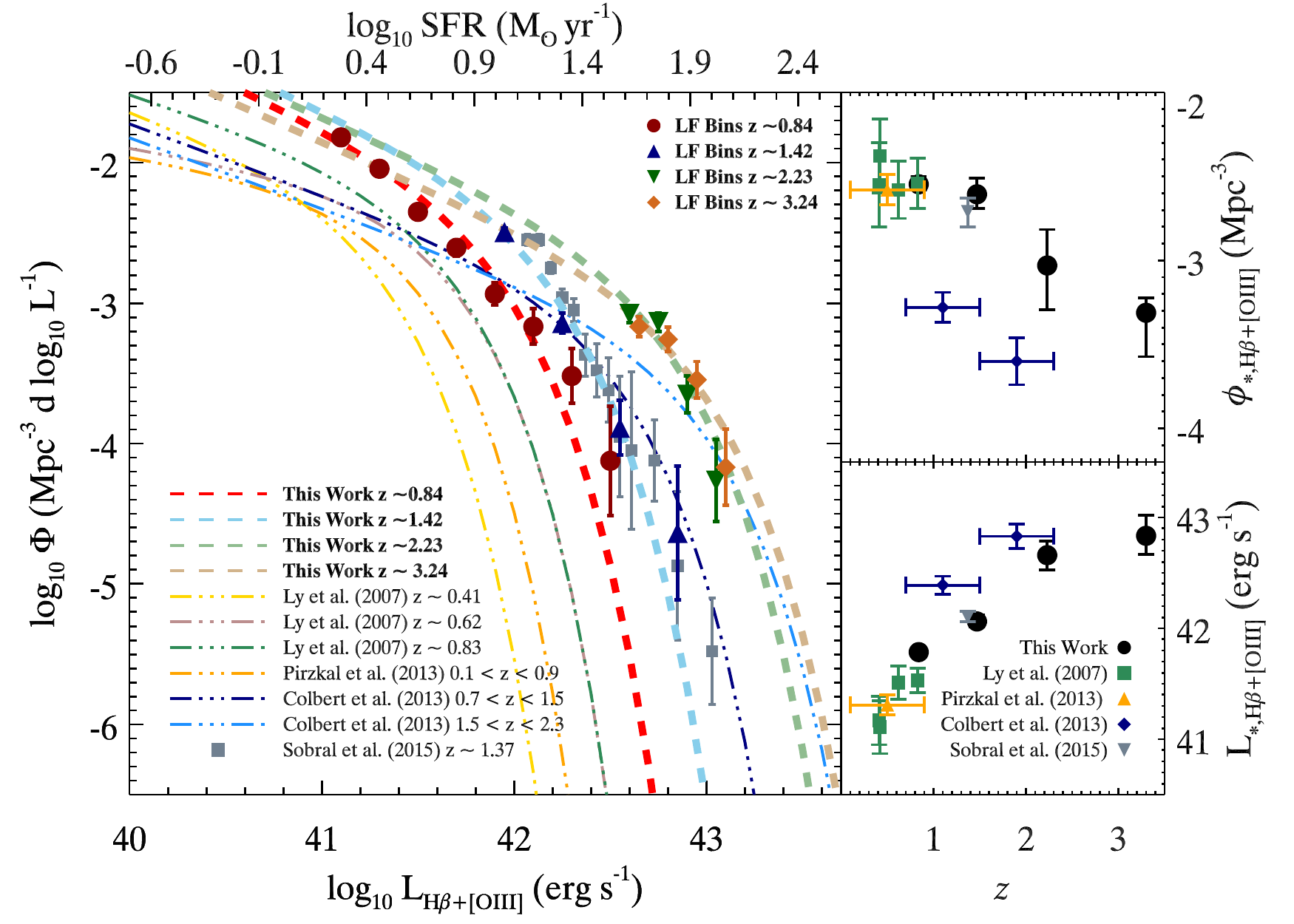}
\caption{{\it Left:} Presented are our \hb~Luminosity Functions along with LFs from the literature. Included on the top horizontal axis is the $\log_{10}$ SFR that was derived via the \citet{Osterbrock2006} calibration (see section \ref{sec:calibrations}). The darker data points are color coded to match the lighter LF fit ({\it dashed lines}). There is a clear evolution in the LFs up to $z \sim3$. We find that our $z = 1.42$ and $2.23$ LFs are in reasonable agreement with the [O\rom{3}] grism spectroscopy study of \citet{Colbert2013} at the bright-end, suggesting that we are selecting a reliable sample of [O\rom{3}] emitters. The major difference between our $z = 0.84$ and the $z = 0.83$ LF of \citet{Ly2007} is probably due to sample size biases. Our sample is much larger, hence we are able to populate our brightest bins, causing a shift in $L_\star$ to higher luminosities. {\it Top Right:} The evolution of $\phi_\star$ from the \hb~luminosity function. A strong, decreasing evolution is seen in $\phi_\star$ from $z = 0$ to $z \sim 3$. This same evolution is seen by UV LF studies (see \citealt{Oesch2010} for details). {\it Bottom Right:} The evolution of $L_\star$  from the \hb~luminosity function. We see a strong, increasing evolution in $L_\star$ up to $z \sim 3$.}
\label{fig:hbeta}
\end{figure*}

We fit each luminosity function using the MPFIT fitting routine \citep{Markwardt2009}, which utilizes the Levenberg-Marquardt least-squares minimization technique to find the best-fit for a given function. For each fit, we take the best-fit values as our fitted parameters. For the 1$\sigma$ errors, we run a Monte Carlo simulation, similar to that of \citet{Ly2011}. The simulation starts by selecting a random number that is drawn from a normal distribution that will perturb each data point, $\Phi(L)$, in the luminosity function within the $1\sigma$ error bars. We also vary the bin size and center of the bin by perturbing the original bin size and bin center by a uniform distribution. The fit is then run again and these steps are repeated for each iteration. A total of $10^5$ iterations are done to get a probability distribution of the best-fit values from where the $1\sigma$ error bars are then calculated. 

Our best-fit values are shown in table \ref{table:lfparams}. We keep $\alpha$ fixed to a constant value of $-1.6$ and $-1.3$ for \hb~ and \oii, respectively, as we are not able to fully constrain the faint-end. These values are drawn from looking at past work from previous narrow-band studies in order for our results to be comparable (e.g., \hb, \citealt{Colbert2013}; \oii, \citealt{Bayliss2011}). The drawback to this is that we are using low-$z$ measurements of $\alpha$ as a proxy for the high-$z$ universe, which can be an incorrect assumption. Low-$z$ studies, such as \citet{Colbert2013}, \citet{Ly2007}, and \citet{Pirzkal2013} for \hb~ and \citet{Ly2007}, \citet{Ciardullo2013}, \citet{Takahashi2007,} \citet{Bayliss2011}, and \citet{Sobral2012} for \oii, have shown that the faint-end slope doesn't evolve up to $z \sim 1.5$, while H$\alpha$ surveys such as \citet{Sobral2013, Sobral2014} have shown no evolution in the faint-end slope up to $z\sim2.23$. Furthermore, UV studies (e.g., \citealt{Oesch2010, Smit2012}), have shown no evolution up to $z \sim 6 - 7$. Based on these results, we keep $\alpha$ fixed as we constrain the bright-end rather than the faint-end. 

Lastly, the LF results are not corrected for dust extinction, except when measuring the star-formation rate densities. This is because many studies in the literature use very different extinction diagnostics, such that it becomes difficult to compare various studies. Furthermore, our knowledge of the role of dust on emission-lines for the high-$z$ universe, especially for the emission-lines of interest in our study, is still in development and requires future detailed investigations. To simplify the use of our LFs by others in future studies, we present all the LF parameter results as uncorrected for dust and AGN contribution. When discussing the SFRDs in section \ref{sec:SFRD}, we will include the results with and without dust and AGN corrections.

\subsection{H$\beta + $[O\rom{3}] Luminosity Function $z \sim $ 0.8, 1.5, 2.2, and 3.3}
\label{sec:hbeta_LF}
We present here the results of the fitted Schechter function to the \hb~observed luminosity function out to $z \sim 3.3$. This is the highest redshift determination of the \hb~LFs currently to date and is the first time that the luminosity function has been constrained out to these redshifts. We present the results in figure \ref{fig:hbeta}. From $z \sim 0.8$ to $3.3$, we see a clear evolution in the shape of the LFs (figure \ref{fig:hbeta}, {\it left}). We also show on figure \ref{fig:hbeta} the evolution of $\phi_\star$ and $L_\star$ with $\alpha$ fixed to $-1.6$. It should be noted that there is a degeneracy between the fitted Schechter parameters, as shown in figure \ref{fig:contours}. This needs to be borne in mind when interpreting the evolution of any single parameter, although figure \ref{fig:contours} indicates that our results are relatively robust. Based on our results, we find that $\phi_\star$ has been decreasing from $z \sim 0.8$ to $\sim 3.3$. The opposite trend is seen in $L_\star$ where it is increasing from $z\sim 0.8$ to $\sim 3.3$.

Our results show a clear evolution in the luminosity function and are consistent with the same evolution seen in the results from the literature \citep{Ly2007,Pirzkal2013,Colbert2013,Sobral2015}. We report our LF parameters in table \ref{table:lfparams}. We find that our $z\sim1.47$ and 2.23 LF agrees well with the LFs of \citet{Colbert2013} in the bright-end, but diverges at the faint-end. This matches with our discussion in the section below (see section \ref{sec:predict}) where we predict the bright-end to be dominated by [O\rom{3}] emitters. The \citet{Colbert2013} study was part of the {\it HST} WISP program, covering 29 fields (0.036 deg$^{2}$) in search of H$\alpha$, [O\rom{3}], and \oii~emission line galaxies using WFC3 grim spectroscopy. Because this was a spectroscopic study and the fact that our LF matches (in the bright-end) with that of \citet{Colbert2013} gives us confirmation that we are picking up the [O\rom{3}] emitters in our sample. The rise in the faint-end can then be attributed to the H$\beta$ emitters in our sample.

In comparison to \citet{Ly2007}, we see a clear deviation of the fits between our $z \sim 0.84$ and their $z\sim0.83$. Although we do agree in terms of $\phi_\star$, the main deviation in the LFs are in $L_\star$ and $\alpha$ (was $\alpha =-1.44\pm0.09$ in comparison to our $-1.6$ fixed faint-end slope). The study was based on deep optical imaging of the Subaru Deep Field (SDF) using the Suprime-Cam on the 8.2 m Subaru Telescope and was complemented with Subaru FOCAS and Keck DEIMOS spectroscopy. The deviation could be due to biases from sample sizes such that our sample consists of more bright emitters to populate the bright-end, hence shifting $L_\star$ higher. In comparison to \citet{Sobral2015}, for which this work was done in unison with, we find perfect agreement but our sample probes deeper by 0.1 dex.

Figure \ref{fig:hbeta} shows the evolution of $L_\star$ along with the results from other studies. There is a strong trend in which $L_\star$ is increasing from $z = 0 - 2.23$ and then flattens. This trend is supported by \citet{Ly2007}, \citet{Pirzkal2013}, \citet{Colbert2013}, and \citet{Sobral2015}. Prior to this work, the $z < 1$ studies hinted to a rising trend in $L_\star$, which with our measurements and the $z \sim 1.5$ measurements of \citet{Sobral2015} has been confirmed up to $z \sim 3$.

For the normalization of the LF, we see an evolution (figure \ref{fig:hbeta}) such that $\phi_\star$ drops as redshift increase up to $z \sim 3$. This is consistent with the collection of UV LFs (i.e., \citealt{Oesch2010}), while our determination is based on a reliable \hb~sample. Note that prior to this study, the \hb~measurements in the literature paint the picture that the $\phi_\star$ evolution is flat up to $z \sim 1$. With the inclusion of our measurements, along with the $z \sim 1.5$ measurements of \citet{Sobral2015}, we find that $\phi_\star$ strong decreases after $z > 1$. 

\begin{figure}
\hspace{-0.7cm}
\includegraphics[scale=0.9]{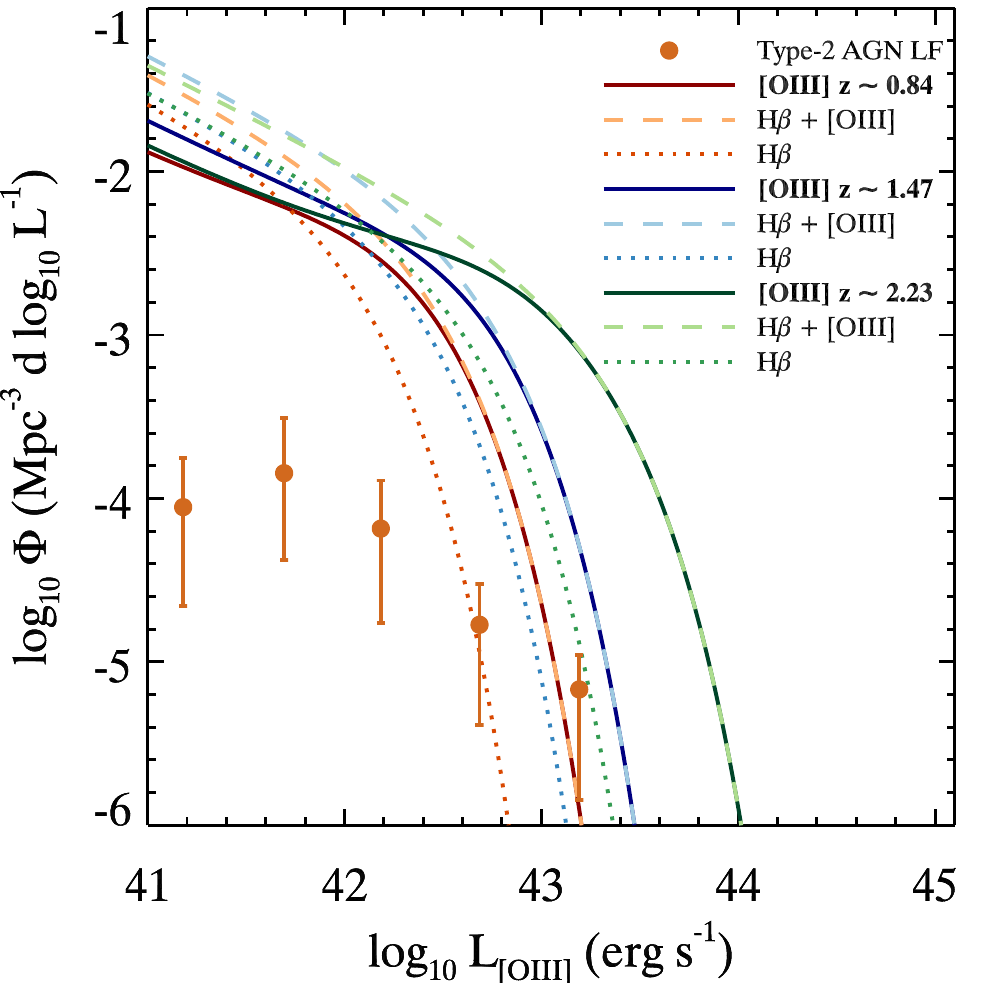}
\caption{Shown is the predicted [O\rom{3}] LFs from $z \sim 0.8$ to 2.2 and compared to the $z\sim 0.71$ zCOSMOS Type-2 AGN LF of \citet{Bongiorno2010}. The H$\beta$ LFs are made by simply taking the H$\alpha$ LFs of \citet{Sobral2013} and assuming a fixed H$\beta$/H$\alpha$ ratio to convert them. We find that our $z\sim0.84$ LF is [OIII]-dominated at the bright-end. Also, the level of AGN contribution is very little, except for $\log_{10}$ L$_{\mathrm{[O\rom{3}]}} > 43.0$ erg s$^{-1}$.}
\label{fig:agn}
\end{figure}  

\begin{figure*}
\centering
\includegraphics{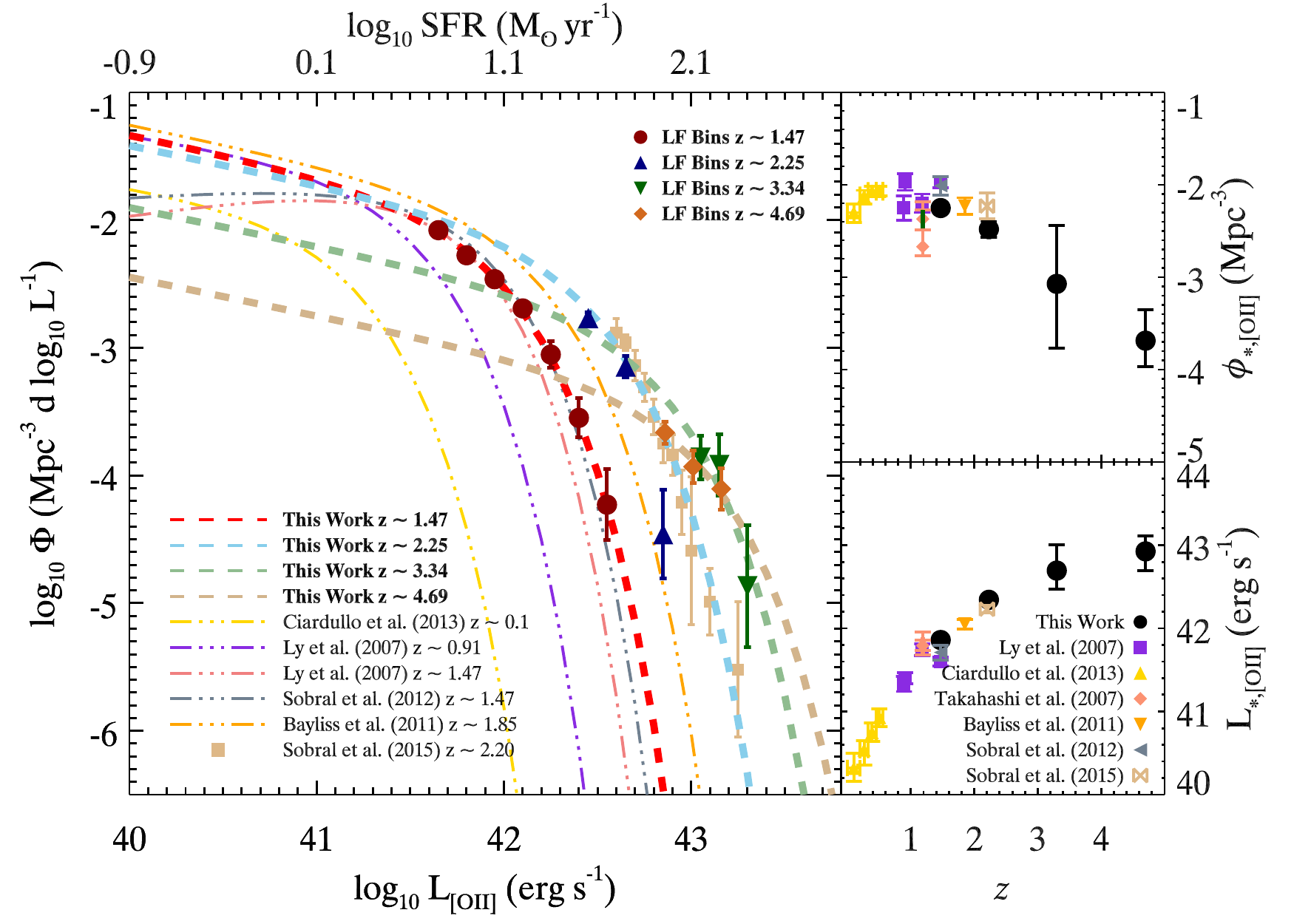}
\caption{{\it Left:} Presented are the \oii~Luminosity Functions along with those from the literature. The SFR calibration used to create the top horizontal axis is from \citet{Kennicutt1998} (see section \ref{sec:calibrations}). The darker data points are color coded to match the lighter LF fit ({\it dashed lines}). We find that the evolution from the low-$z$ studies of \citet{Gallego2002} and \citet{Ciardullo2013} to our $z = 4.7$ LFs is quite strong and clear. We find that our $z = 1.47$ LF is in agreement with the HiZELS \oii~study of \citet{Sobral2012} and the Subaru Deep Survey study of \citet{Ly2007}. Our $z = 2.23$ is also in agreement with the CF-HiZELS study of \citet{Sobral2015}. {\it Top Right:} The evolution in the normalization of the LF. We find that $\phi_\star$ has been decreasing from $z \sim 1.47$ to $z \sim 5$. {\it Bottom Right:} The evolution of $L_\star$. We find a clear, strong evolution in $L_\star$ all the way to $z \sim 5$.}
\label{fig:oii}
\end{figure*}

\subsubsection{Predicting the [O\rom{3}] LF and AGN contribution}
\label{sec:predict}
The results highlighted above are for the H$\beta$ and [O\rom{3}] emitters combined as one sample since we can not separate the two types of emitters based on photometry. For this, we need to conduct spectroscopic follow-ups to properly segregate the emitters. We attempted to separate the sample by using the H$\alpha$ LF of \citet{Sobral2013}. The advantage of using the LFs of \citet{Sobral2013} is that it is fully compatible since we are both using the same data set and methodology. We start by first removing $A_{\mathrm{H}\alpha} = 1.0$ mag dust correction, then apply an H$\beta$/H$\alpha = 0.35$ line ratio from \citet{Osterbrock2006} to get the observed H$\beta$ LF. We then applied $A_{\mathrm{H}\beta} = 1.38$ mag (based on \citet{Calzetti2000}; assuming $A_{\mathrm{H}\alpha} = 1.0$ mag) to the LFs and dust-corrected our \hb~LFs using $A_{\mathrm{H}\beta+\mathrm{[O\rom{3}]}} = 1.35$ (see section \ref{sec:dust} for details). The next step was subtracting our \hb~LFs from the predicted H$\beta$ LFs to get the predicted luminosity function for [O\rom{3}]5007 emitters. The results are shown in figure \ref{fig:agn}.

We find that the [O\rom{3}] emitters in our sample completely dominate the \hb~LFs while towards the faint-end the H$\beta$ emitters dominate. This is expected as the theoretical [O\rom{3}]/H$\beta$ line ratio is $\sim 3$ (for $Z = 0.0004$; \citealt{Osterbrock2006}), which would segregate our sample such that the bright-end will be populated by [O\rom{3}] emitters and the faint-end with H$\beta$ emitters. We also find an interesting feature where the normalization in the [O\rom{3}] LFs are the same, with the exemption of the $z \sim 1.42$ [O\rom{3}] LF which is slightly higher.  This can imply that the relative contribution of H$\beta$ is the same for all three LFs. We note that this is a qualitative assessment and subtracting a Schechter function by another Schechter function doesn't result in the same functional form. 

We also attempted to compare the $z \sim 0.8$ [O\rom{3}] LF with the zCOSMOS AGN Type-2 LF of \citet{Bongiorno2010} to qualitatively assess the contribution of AGNs. The Type-2 AGN is the best candidate that could contaminate our sample as they have a continuum that is similar to normal star-forming galaxies and they photo-ionize the same cold gas that is photo-ionized by hot massive stars. We find that we are in agreement \emph{only} for the brightest luminosity bin of \citet{Bongiorno2010} ($\log_{10}$ L$_{\rm{[OIII]}} \sim 43~ \rm{erg}~ \rm{s}^{-1}$), but in disagreement for the lower luminosity bins. This implies that our brightest [O\rom{3}] emitters are primarily AGNs, but the fainter emitters are a combination of star-forming galaxies and AGNs, with the star-forming galaxies being the most dominant. Future spectroscopic follow-ups of our sample would allow us to properly study the evolution of AGNs in the Universe.

\subsection{[O\rom{2}] Luminosity Function $z = $ 1.47. 2.23, 3.3, and 4.7}
\begin{figure*}
\centering
\includegraphics[scale=0.85]{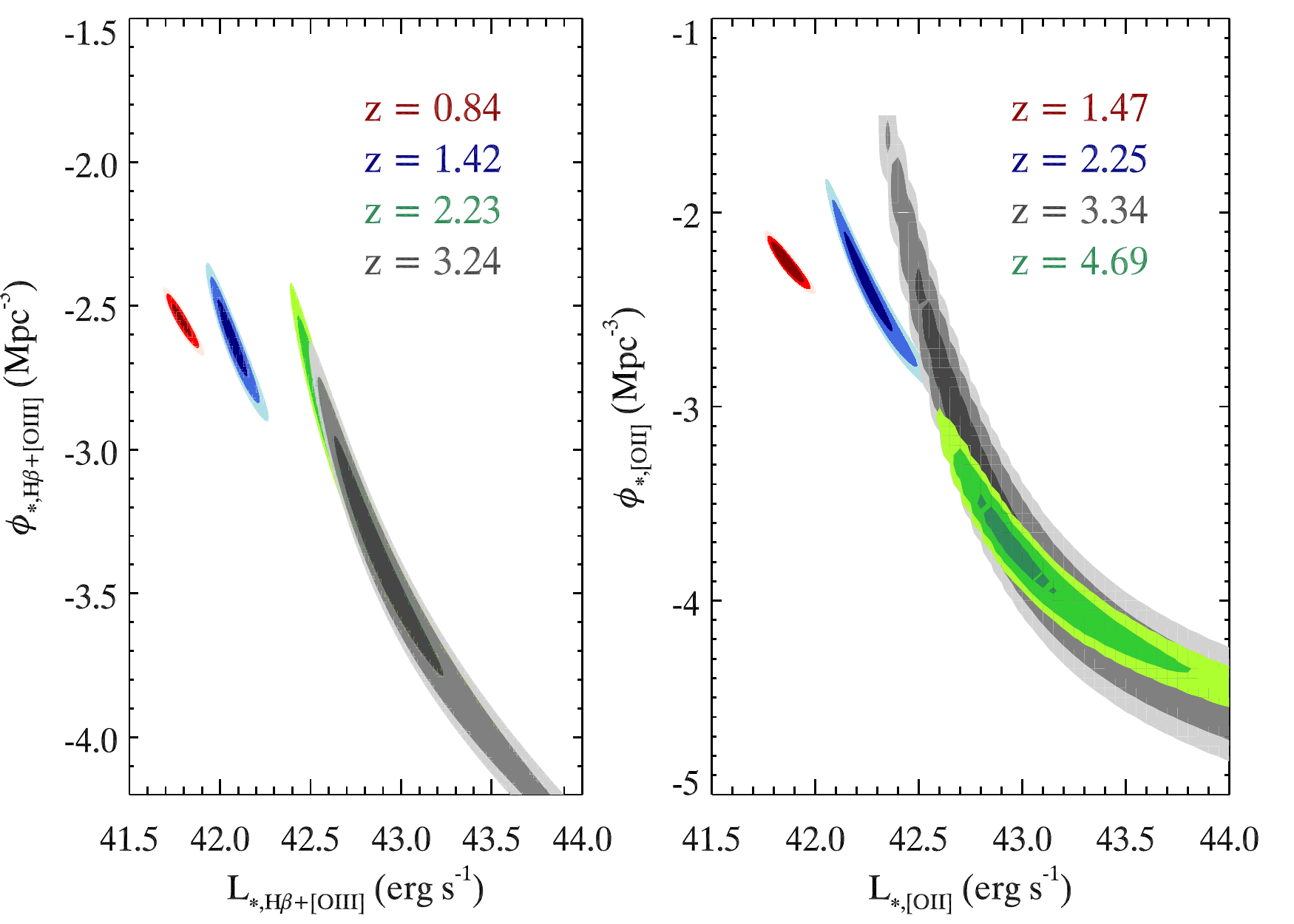}
\caption{Shown is the interdependent evolution of $L_\star$ and $\phi_\star$ for \hb~({\it left}) and \oii~({\it right}). The contours are color-coded to match the text color in the legend. The confidence levels are organized such that the \emph{darkest shade} is the $1\sigma$ level and the \emph{lightest shade} being the $3\sigma$ level. There is a clear evolution that as redshift increases, $\phi_\star$ drops and $L_\star$ increases up to $z\sim 3$ and $z\sim5$ for \hb and \oii, respectively.}
\label{fig:contours}
\end{figure*}

We present here the results of the [O\rom{2}] luminosity function and the Schechter fit out to $z\sim 5$. The results are highlighted in figure \ref{fig:oii}. We see a clear evolution of the LF with redshift, with a large increase in the characteristic luminosity with redshift. The right-hand panels of figure \ref{fig:oii} show the evolution of the fitted $\phi_\star$ and $L_\star$ parameters, and figure \ref{fig:contours} shows the degeneracy between the fitted values.

Included on figure \ref{fig:oii} are data from the literature that range from $z = 0-2.2$ \citep{Bayliss2011, Ciardullo2013, Ly2007, Takahashi2007, Sobral2012, Sobral2015}. We find that our $z = 2.23$ binned LF data is in agreement with the CF-HiZELS result of \citet{Sobral2015}. Because their sample size is $\sim 4$ times larger than our measurement, we have combined their LF data points with ours. The main effect is our measurement extends the combined LF 0.15 dex fainter. We note that the LFs of \citet{Sobral2015} are directly compatible with our LFs as our study follows the same methodology. In fact, the \citet{Sobral2015} is specific to the NBJ determined LFs and the effects of cosmic variance while this study focuses on the evolution and extension of the LFs out to $z \sim 5$. 

Our $z = 1.47$ measurements are in perfect agreement with \citet{Sobral2012} and close to agreement with \citet{Ly2007}. We note that for \citet{Ly2007} the faint-end was measured to be $\alpha=-0.78\pm0.13$ and $-0.9\pm0.2$ for \citet{Sobral2012} while we keep our faint-end slope fixed to $-1.3$ for all LFs. As seen in figure \ref{fig:oii}, we find that we are not probing deep enough to fully see the turn in the LF for $L_\mathrm{[O\rom{2}]} < 41 $ erg s$^{-1}$ as found by \citet{Ly2007} and \citet{Sobral2012}. This is probably due to the fact that both studies used $2''$ apertures, while our study uses $3''$ apertures (provided in the \citet{Sobral2013} catalog) to select $z = 1.47$ \oii~emitters, which means that their studies are better at recovering faint emitters. 

We find a strong evolution in $L_\star$, as shown in figure \ref{fig:oii} for which a strong, rising trend is seen up to $z \sim 5$. We also see the same evolution in H$\alpha$ studies (e.g., \citet{Sobral2012}), where $L_\star$ is strongly increasing from $z = 0$ to 2. The same evolution of $L_\star$ is seen in UV studies \citep{Oesch2010} up to the same redshift range. We notice some scatter for the low-$z$ studies (\citealt{Ly2007,Takahashi2007,Bayliss2011,Sobral2012,Ciardullo2013}). This is primarily due to limitations in survey area and/or shallowness of the surveys.

%- LF PARAMS TABLE
\begin{table*}
\centering
\caption{The Luminosity Function parameters and derived properties for all \hb~and \oii~emitters. Errors in $\phi_\star$ and $L_\star$ are computed by running a Monte Carlo Simulation, displacing all the measurements of $\Phi_{\mathrm{final}}$ by $1\sigma$. The faint-end slope, $\alpha$, was fixed as our data don't go faint enough to constrain the faint-end properly. The luminosity density, $\rho_L$, was calculated by taking the infinite integral of the LF. The SFRDs were calculated based on the \citet{Kennicutt1998} calibration (\oii) and \citet{Osterbrock2006} (\hb) with $\dot{\rho}_{\star, \rm comp}$ being the completeness and filter profile corrected SFRD measurement. $\dot{\rho}_{\star, \rm corr}$ is the completeness + filter profile + dust corrected SFRD measurement. $\dot{\rho}_{\star, \rm AGN-corr}$ is the completeness + filter profile + dust corrected + AGN corrected SFRD measurement. We show every measurement as $\dot{\rho}_{\star, \rm comp}$ is the most robust measurement, while the dust and dust + AGN corrected measurements are based on assumptions regarding line ratios, dust extinction laws, and AGN selection methods.}
\begin{tabular}{c c c c c c c c}
\hline
\multicolumn{8}{c}{H$\beta +$ [O\rom{3}] Luminosity Function Properties}\\
\hline
\hline
$z$ & $\log_{10} \phi_\star$ & $\log_{10} L_\star$ & $\alpha$ & $\log_{10} \rho_L$  & $\log_{10} \dot{\rho}_{\star, \rm comp}$ & $\log_{10} \dot{\rho}_{\star,\rm corr}$ & $\log_{10} \dot{\rho}_{\star, \rm AGN-corr}$\\
 & (Mpc$^{-3}$) & (ergs s$^{-1}$) &  & (ergs s$^{-1}$ Mpc$^{-3}$) & (M$_\odot$ yr$^{-1}$ Mpc$^{-3}$) & (M$_\odot$ yr$^{-1}$ Mpc$^{-3}$) & (M$_\odot$ yr$^{-1}$ Mpc$^{-3}$)\\
\hline
$0.84$ & $-2.55^{+0.04}_{-0.03}$ & $41.79^{+0.03}_{-0.05}$ & $-1.60$ & $39.58$ & $-1.549^{+0.01}_{-0.02} $&$ -1.009^{+0.01}_{-0.02} $&$ -1.062^{+0.03}_{-0.03}$ \\
$1.42$ & $-2.61^{+0.10}_{-0.09}$ & $42.06^{+0.06}_{-0.05}$ & $-1.60$ & $39.80$ & $-1.333^{+0.05}_{-0.04} $&$ -0.793^{+0.05}_{-0.04} $&$ -0.882^{+0.06}_{-0.05}$ \\
$2.23$ & $-3.03^{+0.21}_{-0.26}$ & $42.66^{+0.13}_{-0.13}$ & $-1.60$ & $39.98$ & $-1.159^{+0.10}_{-0.11} $&$ -0.619^{+0.10}_{-0.11} $&$ -0.766^{+0.11}_{-0.12}$ \\
$3.24$ & $-3.31^{+0.09}_{-0.26}$ & $42.83^{+0.19}_{-0.17}$ & $-1.60$ & $39.87$ & $-1.265^{+0.10}_{-0.09} $&$ -0.725^{+0.10}_{-0.09} $&$ -0.873^{+0.11}_{-0.10}$ \\
\hline
\multicolumn{8}{c}{\oii~ Luminosity Function Properties}\\
\hline
\hline
$z$ & $\log_{10} \phi_\star$ & $\log_{10} L_\star$ & $\alpha$ & $\log_{10} \rho_L$ & $\log_{10} \dot{\rho}_{\star, \rm comp}$ & $\log_{10} \dot{\rho}_{\star, \rm corr}$ & $\log_{10} \dot{\rho}_{\star, \rm AGN-corr}$\\
 & (Mpc$^{-3}$) & (ergs s$^{-1}$) &  & (ergs s$^{-1}$ Mpc$^{-3}$) & (M$_\odot$ yr$^{-1}$ Mpc$^{-3}$) & (M$_\odot$ yr$^{-1}$ Mpc$^{-3}$) & (M$_\odot$ yr$^{-1}$ Mpc$^{-3}$)\\
\hline
$1.47$ & $-2.25^{+0.04}_{-0.04}$ & $41.86^{+0.03}_{-0.03}$ & $-1.30$ & $39.72$ & $-1.132^{+0.02}_{-0.02} $&$ -0.884^{+0.02}_{-0.02} $&$ -0.973^{+0.04}_{-0.04}$ \\
$2.25$ & $-2.48^{+0.08}_{-0.09}$ & $42.34^{+0.04}_{-0.03}$ & $-1.30$ & $39.98$ & $-0.878^{+0.05}_{-0.06} $&$ -0.630^{+0.05}_{-0.06} $&$ -0.723^{+0.06}_{-0.07}$ \\
$3.34$ & $-3.07^{+0.63}_{-0.70}$ & $42.69^{+0.31}_{-0.23}$ & $-1.30$ & $39.74$ & $-1.118^{+0.43}_{-0.20} $&$ -0.870^{+0.43}_{-0.20} $&$ -0.964^{+0.43}_{-0.20}$ \\
$4.69$ & $-3.69^{+0.33}_{-0.28}$ & $42.93^{+0.18}_{-0.24}$ & $-1.30$ & $39.35$ & $-1.502^{+0.10}_{-0.10} $&$ -1.255^{+0.10}_{-0.10} $&$ -1.348^{+0.11}_{-0.11}$\\
\hline
\hline
\end{tabular}
\label{table:lfparams}
\end{table*}

The evolution in $\phi_\star$ is shown in figure \ref{fig:oii} along with measurements from the literature. Our measurements show a decreasing trend since $z \sim 1.5$ while for redshifts less than 1.5 shows a flat evolution in $\phi_\star$.The same evolution in $\phi_\star$ is also seen in H$\alpha$ studies (e.g., \citet{Sobral2013}) where after $z \sim 1$ and up to $z \sim 2$, $\phi_\star$ is shown to be decreasing. UV measurements \citep{Oesch2010} also see a similar trend. 

Future wide surveys, such as {\it Euclid} and WFIRST, will be able to observe larger samples of emission-line galaxies such that our results can be used as predictions for such upcoming projects. By taking our LFs and integrating them to some flux/luminosity limit, these future surveys can estimate the number of \oii~emitters that can be detectable. Our LFs would then be quite useful as a tool to plan surveys studying \oii~emitters out to $z \sim 5$. Furthermore, our luminosity functions can be used as a tool to gauge the level of low-$z$ interlopers in various studies, such as Ly$\alpha$ studies at high-$z$.

\section{Evolution of the Star-Formation History of the Universe}
\label{sec:SFRD}
In this section, we present the star-formation history evolution of our \oii~sample out to $z \sim 5$.  We begin by measuring the level of AGN contamination and then present the calibrations used to get the star-formation rate densities (SFRDs). We conclude with a discussion of the evolution of the SFRD based on \oii~emitters, the correction for dust, and our estimates of the stellar mass density evolution of the universe based on our SFRD fit.

\subsection{Contribution from AGN}
\label{sec:agn}
Active Galactic Nuclei (AGN) play an important role in the evolution of galaxies. Because AGN heat the cold gas that is photo-ionized by O \& B-type stars in star-forming regions, the same emission-lines become present by both sources. It is then imperative that the SFRDs are properly corrected for AGN contamination to ensure that the sample is, by majority, a star-forming sample. Due to the low number-density of AGNs, it is difficult to use the current catalogs in the literature (e.g., {\it Chandra-COSMOS}) as a direct indicator on the level of AGN contribution/contamination to our sample. When comparing to {\it Chandra-COSMOS} \citep{Elvis2009}, we find 1, 0, 4, 0 for $z \sim 0.84$, 1.42, 2.23, and 3.24 for our \hb~sample and 5, 2, 1, 0 for $z \sim 1.47$, 2.25, 3.34, and 4.69 for our \oii~sample. We also compared our catalogs to the {\it XMM-COSMOS} catalog \citep{Cappelluti2009} and found no matches.

These matches themselves can not give us a complete indication of the level of our AGN contamination as they are X-ray flux-limited. We instead take advantage of the rest-frame 1.6\micron~bump in the SEDs of star-forming galaxies. This bump arises from the minimum opacity of H$^{-}$ ions in the stellar atmospheres of cool stars. For AGNs, the bump is meshed in with various other emission (e.g., PAHs, silicate grains) resulting in a rising power-law SED after 1.6\micron~in the rest-frame. We use the deep IRAC data in COSMOS and UDS and with the condition that redder colors are AGNs, signifying the rising SED after 1.6\micron, and anything with bluer colors are star-forming galaxies resulting in the 1.6\micron~bump. We measure the colors by taking the $[3.6 - 4.5] > 0.1$ ($z \sim 0.8$), $[4.5 - 5.6] > 0.1$ ($z \sim 1.5$), and $[5.6 -8.0] > 0.1$ ($z \sim 2.2$) for \hb~emitters and $[4.5 - 5.6] > 0.1$ ($z \sim 1.47$) and $[5.6 - 8.0] > 0.1$ ($z \sim 2.23$) for \oii~emitters. We find AGN contamination for \hb~ is $\sim 11.4\%$, $\sim 18.5\%$, and $\sim 28.8\%$ for $z \sim 0.8$, $1.5$, and $2.2$, respectively. The amount of AGN contamination for \oii~is $\sim18.5\%$ and $\sim19.4\%$ for $z\sim 1.47$ and $2.23$, respectively. For $z > 2.2$ in \oii~emitters, we set the AGN contamination constant to that at $z = 2.2$ as this would require going beyond the last IRAC band. Note that these are upper limits for the level of AGN contamination such that our SFRDs are corrected for the highest contamination possible via the 1.6\micron~bump technique. By comparing our SFRD measurements to other star-formation tracers, we can determine if the AGN correction was too high or not. But to reliably measure the correction will require follow-up spectroscopy of our sample to properly separate the AGN from the star-forming sample. We therefore apply our determined AGN correction to the luminosity densities measured from the fully integrated LFs (decreases the SFRDs) and include 20\% of the correction factor in quadrature with the luminosity density errors.

\subsection{Calibrations}
\label{sec:calibrations}
\begin{figure}
\hspace{-1.5cm}
\includegraphics{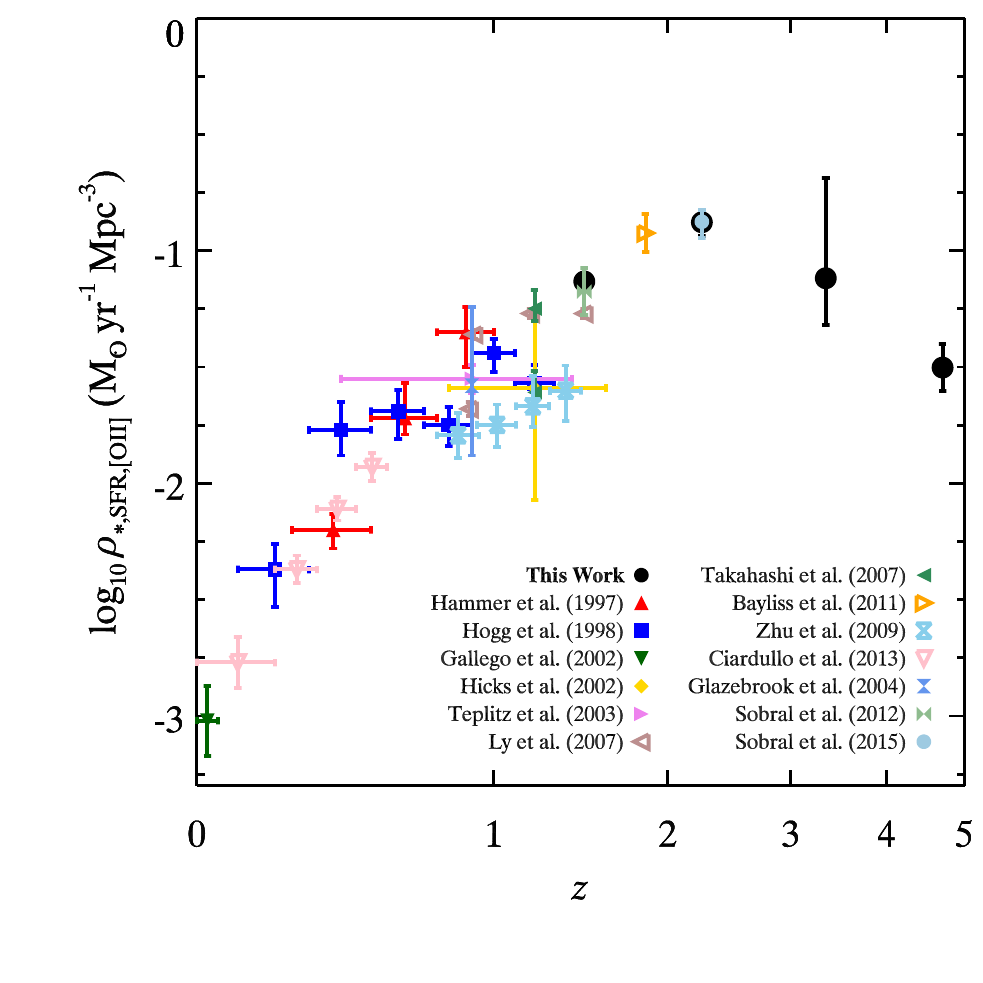}
\caption{The uncorrected for dust SFRD evolution based only on \oii~emission studies. We find that our $z = 1.47$ and $2.25$ LF continues the inverse power-law slope that is found from $z = 0 - 2$ in the majority of SFRD studies, and a continuous drop for $z > 2$. This is the first time that \oii~studies have gone beyond $z \sim 1.5$ in a reasonably, statistically constrained fashion. We find perfect agreement with \citet{Sobral2012} and are 0.15 dex off from \citet{Ly2007}. Also, we find perfect agreement with the $z = 2.25$ SFRD of \citet{Sobral2015}.}
\label{fig:sfrd}
\end{figure}

The star-formation rate density is calculated via the luminosity density from the LF at each redshift. The luminosity density is defined as:
\begin{equation}
\Ld = \int_{0}^\infty \Phi(L) L \mathrm{d}L = \phi_\star L_\star \Gamma(\alpha+2)
\label{eqn:ld}
\end{equation}
where $\Ld$ is the luminosity density, $\phi_\star$ is the normalization, $L_\star$ is the characteristic luminosity, and $\alpha$ is the faint-end slope. Our determined luminosity densities are highlighted in table \ref{table:lfparams} and consider the full range of luminosities. The star-formation rate density (SFRD) is then calculated by using the \citet{Kennicutt1998} diagnostics:
\begin{equation}
\dot{\rho}_\mathrm{SFR,O\rom{2}} = 1.4\e{-41} \Ld_{\mathrm{[O\rom{2}]}} M_\odot \mathrm{yr}^{-1} \mathrm{Mpc}^{-3}
\label{eqn:calib_OII}
\end{equation}
where a $\Ld_{\mathrm{[O\rom{2}]}}/\Ld_{\mathrm{H}\alpha} = 1.77$ is assumed. We note that using the \oii~SFR calibration comes with several drawbacks, such as metallicity, reddening, and line ratio assumptions, but the \oii~line is the brightest emission-line detectable at $z > 1.5$ where H$\alpha$ falls in the infrared. We will present the effects of these drawbacks in a future study (Khostovan et al. in prep). Furthermore, we present the uncorrected for dust SFRD measurements to see, qualitatively, the evolution of the SFRD. This means that the results shown in figure \ref{fig:sfrd} are lower-limits since any dust correction will just increase the SFRD measurements.

We also use the derived relation of \citet{Osterbrock2006}:
\begin{equation}
\dot{\rho}_\mathrm{SFR,H\beta+O\rom{3}} = 7.35\e{-42} \Ld_{\mathrm{[H\beta+O\rom{3}]}} M_\odot \mathrm{yr}^{-1} \mathrm{Mpc}^{-3}
\label{eqn:calib_HB}
\end{equation}  
to measure the \hb~SFRD\footnotemark~although this can not be taken as a purely star-forming indicator as there are several caveats behind it. \emph{We want to make this point specifically clear; we do not use the \hb~SFRDs in fitting the star-formation and stellar mass assembly history of the Universe}. We instead use it to compare the measurements to those of the \oii~SFRDs and other tracers in the literature to show if our sample is tracing a star-forming sample and whether or not if the \hb~calibration is more of a ``reliable'' tracer of star-formation activity than previously thought.

\footnotetext{We used the dust extinction curve of \citet{Calzetti2000} for the \hb~emitters and applied for all redshifts, such that $A_{\mathrm{H}\beta+\mathrm{[O\rom{3}]}} = 1.35$ mag (assuming $A_{\mathrm{H}\alpha} = 1.0$ mag).}

\subsection{Star-Formation Rate Density Evolution of \oii~Emitters}
\subsubsection{Uncorrected-for-Dust}
Figure \ref{fig:sfrd} shows the evolution of the uncorrected-for-dust [O\rom{2}] SFR density for the first time and determined in a self-consistent way from $z = 0 - 5$. The evolution is clear and signifies that a peak that occurs at $z \sim 2 - 3$ and then there is a fall for higher redshifts. We include \oii~measurements from the literature after uncorrecting them for dust and correcting the cosmology (the pre-2000 papers used non-$\Lambda$CDM cosmological parameters). These results plus ours can then be taken as a lower limit as any dust extinction correction would increase the SFR densities. We include a compilation of SFR densities and LF parameters from various studies, using various diagnostics, all normalized to the same cosmology as that of this paper, same \citet{Kennicutt1998} calibration with the same line ratio, and all with $A_{\mathrm{[O\rom{2}]}} = 0$ to make it easier for future studies to utilize. This compilation is found in appendix \ref{sec:sfrd_comp}. We find that our $z = 1.47$ \oii~SFRD measurement is in perfect agreement with \citet{Sobral2012} and \citet{Ly2007}. We are also in perfect agreement with the measurement of \citet{Sobral2015} and our $z = 2.23$ measurement.
  
We note that \citet{Bayliss2012} made a measurement at $z = 4.6$ by observing the GOODS-S field using the NB2090 and Ks filters of the ESO HAWK-I instrument. The results of this study are restricted to a sample of only 3 genuine emitters. Although ground-breaking at the time, their SFRD estimate is severely limited by issues of sample size and cosmic variance.

\subsubsection{Dust \& AGN Corrected SFRD}
\label{sec:dust}

To compare with other studies using different SFRD diagnostics, we adopt a dust correction using the HiZELS $z = 1.47$ measurement of \citet{Hayashi2013}, $A_{\mathrm{H}\alpha} \sim 0.35$ mag. \citet{Hayashi2013} studied H$\alpha$ and \oii~emitters using HiZELS data to conclude that the traditional $A_{\mathrm{H}\alpha} = 1.0$ mag that has been used in the literature is overestimating the dust correction for \oii~emitters at $z = 1.47$, such that these emitters are observed to have $A_{\mathrm{H}\alpha} \sim 0.35$ mag and are less dusty than previously thought.

\begin{figure*}
\centering
\includegraphics{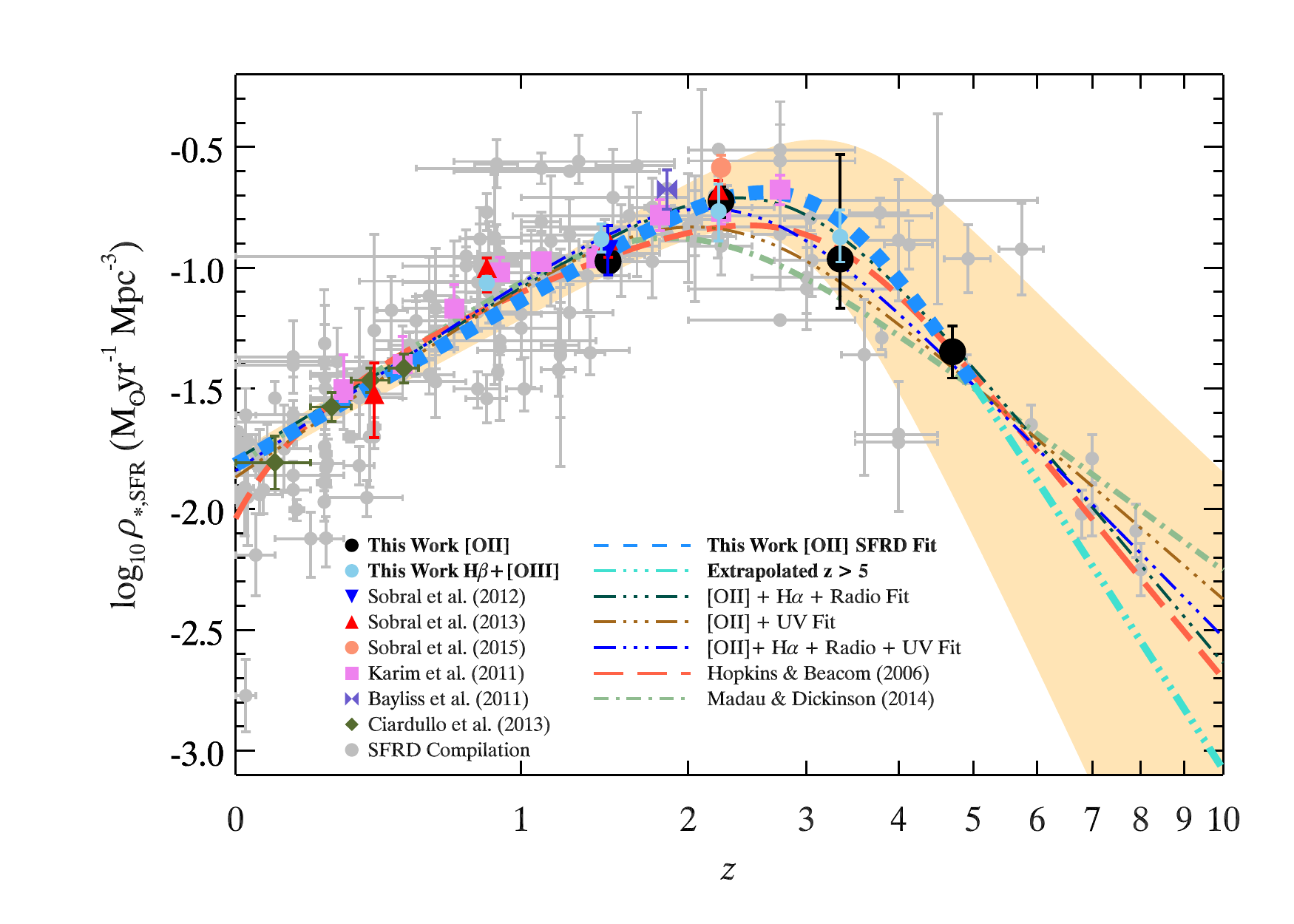}
\caption{Our \oii~dust \& AGN corrected SFRD evolution with the \oii~studies of \citet{Bayliss2011,Ciardullo2013, Sobral2013} and \citet{Sobral2015}, along with the results of this paper, that are used to fit the parametrization of \citet{Madau2014}. The best fit is shown as the dashed line ({\it dodger blue}) and is only based on \oii~measurements. We also include an extrapolation to higher-$z$ ({\it dashed-dotted turquoise} line), as we don't constrain this part of redshift space but can extrapolate based on our fit. The 1-$\sigma$ region is highlighted in {\it gold} filled regions around the fit. The stacked radio study of \citet{Karim2011} and the H$\alpha$ study of \citet{Sobral2013} are also shown as a comparison and are in agreement with our measurements. Our compilation of SFRD measurements (in {\it gray}) are a combination of our compilation and that of \citet{Hopkins2006}, \citet{Madau2014}, \citet{Ly2007}, and \citet{Gunawardhana2013}. We reproduce the SFRD evolution history of the universe based primarily on \oii~studies with the peak of star-formation history occurring at $z \sim 3$. We also include the fits of \citet{Hopkins2006} (IMF corrected to Salpeter) and that of \citet{Madau2014}. We find that the \citet{Hopkins2006} fit reasonably matches our SFRD fit, while the \citet{Madau2014} fits well until $z > 2$. This is mostly because the \citet{Madau2014} fit is driven by the $z > 5$ UV measurements (which are not backed by spectroscopy), for which we do not include in our \oii~fit.}
\label{fig:sfrd_full}
\end{figure*}

To test the dust extinction coefficient of \citet{Hayashi2013}, we apply the traditional $A_{\mathrm{H}\alpha} \sim 1.0$ mag to all four \oii~SFRD measurements. We find that based on this dust correction, our measurements overestimate the H$\alpha$-based SFRD measurements of \citet{Sobral2013} and the radio-stacked measurements of \citet{Karim2011}, which is impervious to dust extinction. The level of overestimation is such that our $z = 1.47$ SFRD measurement and $z = 2.23$ SFRD measurement was $\sim 0.4$ dex above the SFRD measurements of \citet{Sobral2013} and \citet{Karim2011}. When using the \citet{Hayashi2013} dust extinction coefficient, we find that our SFRD measurements are perfectly matched with \citet{Sobral2013} and \citet{Karim2011}, as seen in figure \ref{fig:sfrd_full}. 

We apply the Calzetti correction \citep{Calzetti2000} with the \citet{Hayashi2013} measurement of $A_{\rm{H}\alpha} \sim 0.35$ mag such that:
\begin{equation}
\frac{A_\mathrm{\oii}}{A_{\mathrm{H}\alpha}} = \frac{k(\mathrm{\oii})}{k(\mathrm{H}\alpha)}
\label{eqn:dust}
\end{equation}
where $k(\mathrm{\oii}) = 5.86$ and $k(\mathrm{H}\alpha) = 3.31$, resulting in $A_{\mathrm{\oii}} = 0.62$ mag. We calibrate all the measurements to the same \oii~SFR calibration of \citet{Kennicutt1998}. All measurements hereinafter include AGN corrections as discussed in section \ref{sec:agn}.

Figure \ref{fig:sfrd_full} shows our dust-corrected and AGN-corrected \oii~SFRD measurements. We also include a large compilation of studies from the literature which is a combination of the compilations of \citet{Hopkins2006}, \citet{Madau2014}, \citet{Gunawardhana2013}, \citet{Ly2007}, and our own compilation as a comparison (appendix \ref{sec:sfrd_comp}). We find that our measurements accurately reproduce the star-formation history of the universe up to $z \sim 5$. This is the first time that an \oii~study has ever accomplished such a measurement in a self-consistent manner. We find that the $z = 1.47$ and $2.23$ perfectly agree with the HiZELS H$\alpha$ measurements of \citet{Sobral2013} on figure \ref{fig:sfrd_full}. The AGN contamination in \citet{Sobral2013} assumed a simple $\sim10\%$, which is backed by a detailed search of potential AGNs by \citet{Garn2010}. Based on the similarities between our \oii~measurement and the independently AGN-corrected SFRD measurement of \citet{Sobral2013}, we can conclude that the level of AGN contamination measured is reasonable and the methodology sound.

Another key point is the stacked radio measurements of \citet{Karim2011} ({\it pink squares} on figure \ref{fig:sfrd_full}). The benefit of radio measurements are that they are impervious to dust, but have the downside of poor resolution and blending. We find that our $z = 1.47$ and $z = 2.25$ \oii~measurements are in agreement with \citet{Karim2011}, such that the \citet{Hayashi2013} dust extinction coefficient does reliably correct our measurements to represent the dust-corrected SFRD of star-forming galaxies.

We also find an interesting result when comparing the \hb~SFRD measurements to our \oii~SFRDs and other measurements in the literature. As discussed above, the \hb~calibration is considered in the literature as a ``mixed'' tracer of star-formation activity. Here we find that using the calibration of \citet{Osterbrock2006} with a $A_{\rm{H}\beta+\rm{[O\rom{3}]}} = 1.35$ mag (based on the traditional $A_{\rm{H}\alpha} = 1$ mag), our \hb~measurement for $z = 0.84$ matches perfectly with the H$\alpha$ SFRD of \citet{Sobral2013} and the radio measurement of \citet{Karim2011}. The implications of this agreement shows that not only is the dust correction technique applied correctly, but also that the AGN correction is accurate such that it is matching with the AGN-corrected H$\alpha$ SFRD of \citet{Sobral2013}. For our $z = 1.43$ \hb~SFRD, we find a perfect match with \citet{Karim2011}, \citet{Sobral2012}, \citet{Sobral2013}, and our \oii~SFRD. The $z = 2.23$ \hb~SFRD matches well with \citet{Karim2011}, \citet{Sobral2012}, and our \oii~SFRD measurement. Lastly, we find perfect agreement between our \oii~SFRD and the \hb~SFRD at $z \sim 3.3$. All these perfect agreements hint to the notion that the \hb~SFR calibrations could in fact be more of a reliable tracer of star-formation activity than previously thought. Furthermore, this is also strong evidence to show that our \hb~sample is dominated by star-forming galaxies and is a reliable sample. Also, our survey seems to be detecting \hb~emitters that have more dust in comparison to \oii~emitters such that the traditional $A_{\rm{H}\alpha} = 1$ mag applies to the \hb~sample and a lower dust correction applies to the \oii emitters. This notion was proposed by \citet{Hayashi2013} for their \oii~sample. Their conclusion was that dustier \oii~emitters fall to lower luminosities that are below the detection limit, while the less dusty emitters, which will be apparently brighter, are detected.

We fit the SFRD using our \oii~SFRD measurements along with the \oii~measurements of \citet{Bayliss2011, Ciardullo2013} and \citet{Sobral2012} to the parametrization of \citet{Madau2014}:
\begin{equation}
\log_{10} \dot{\rho}_\star = a \frac{(1+z)^b}{1+[(1+z)/c]^d}~\mathrm{M}_\odot ~\mathrm{yr}^{-1}~\mathrm{Mpc}^{-3}
\label{eqn:sfrd}
\end{equation}
where our fit results with $a = 0.015\pm0.002$, $b = 2.26\pm0.20$, $c = 4.07\pm0.51$, and $d = 8.39\pm2.60$. The fit is purely based on \oii~emitters, but we have also fitted for the cases of \oii+H$\alpha$+Radio, \oii+UV, and \oii+H$\alpha$+Radio+UV (see figure \ref{fig:sfrd_full}) to show how our fit will vary based on the data that we use. Based on the \oii~fit, we see a drop at $z >3$ that is slightly steeper than those determined by UV dropout studies (i.e., \citealt{Bouwens2011,Bouwens2014,Oesch2010,Schenker2013}). Despite this drop in our \oii~SFRD compared to the UV studies, we do find that the UV measurements are still within 1$\sigma$.

An important note to make though is that prior to this paper, there does not exist a study besides UV/Ly$\alpha$ studies that have measured the SFRD up to $z \sim 5$ since $z \sim 3$. This is a crucial point since there has been no other study so far that could confirm the drop-out measurements, which are severely affected by dust extinction. Furthermore, this is the first time that the cosmic star-formation history has been constrained based on a single tracer for larger volumes and up to $z \sim 5$. Our current measurements are the farthest that we can measure the \oii~SFRD due to the fact that the emission line would go past $K$-band and into the infrared. Future space-based narrow-band surveys, such as {\it JWST} and the Wide-field Imaging Surveyor for High-redshift ({\it WISH}), will be able to probe \oii~emitters up to $z \sim 12$, which would allow us to compare and confirm the UV SFRD measurements at $z > 5$.

We also compare our fit to those of \citet{Hopkins2006} and \citet{Madau2014} in figure \ref{fig:sfrd_full}. For the $z < 2$ regime, we find that our \oii~SFRD fit agrees well with all the other fits. For the $z > 2$ regime, we do see divergences based on the fit. In terms of the actual data points, we find that the \citet{Hopkins2006} is closest in agreement as it has a continuing SFRD up to a peak at $z\sim 2.5$ and a drop that continues through the high-$z$ \oii~measurements. The \citet{Madau2014} is also in agreement for the high-$z$ measurements, but fails to match with the $z \sim 2 - 3$ peak. This is mostly due to the fact that their measurements are driven by the $z > 5$ UV dropout SFRDs (e.g., \citealt{Bouwens2011, Bouwens2014, Oesch2010, Schenker2013}).

As with all SFR measurements, there are systematic uncertainties that must be taken into account. In the case of \oii~emitters, our main systematic uncertainties come from metallicity and dust extinction. To study the metallicities and its effects on the star-formation rate calibration, we will need to conduct follow-up spectroscopy. Furthermore, studying the metallicity of our sample will give us also an understanding of the dynamics (inflow/outflow) that can affect star-formation activity. We also plan to study in a future paper the dust extinction properties of our sample and how it relates to and affects the star-formation activity of galaxies in our sample (Khostovan et al., in prep).

\subsection{Evolution of the Stellar Mass Density}
We use the \oii~SFRD results presented in this paper to provide an estimate of the stellar mass density (SMD) evolution by doing a time-integral of equation \ref{eqn:sfrd}. The SMD evolution gives us an understanding of how the universe has assembled its mass throughout cosmic time. This estimate is quantitatively sensitive to the choice of the IMF in terms of the normalization of the SFRD and SMD evolution, but it does not qualitatively affect the final results.

Our estimate assumes a Salpeter IMF, which has been used throughout this entire paper, and a recycling fraction of $R = 0.27$. For a review of the derivation of this factor, we refer the reader to the recent review of \citet{Madau2014}. We calculate the SMD by:
\begin{equation}
\rho_\star(z) =  (1 - R) \int_z^\infty \frac{\dot{\rho}_\star(z)}{H_0(1+z) \sqrt{\Omega_M (1+z)^3 + \Omega_\Lambda}} \mathrm{d} z
\end{equation}

where $\dot{\rho}_\star(z)$ is the SFRD fit using the parametrization defined in equation \ref{eqn:sfrd} and $R$ is the recycling fraction, or the fraction of stars that is returned back into the ISM and IGM. Because the equation above is using $z = \infty$ as a reference for which we do not know the SMD for, we instead constrain the integral such that at $z = 0$ the SMD will be $\log_{10} \rho_\star \sim 8.6 ~ \mathrm{M}_\odot ~ \mathrm{Mpc}^{-3}$, in agreement with measurements made at that redshift.

\begin{figure}
\hspace{-1.25cm}
\includegraphics{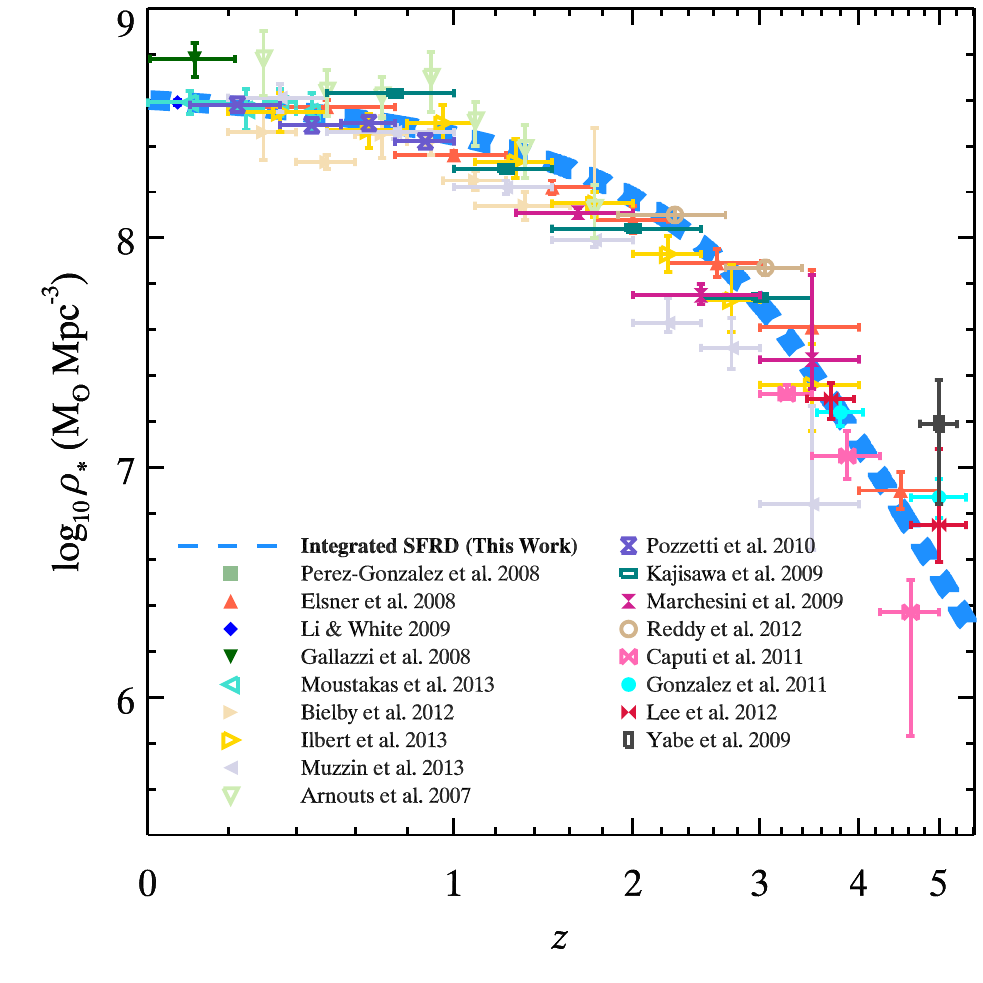}
\caption{The evolution of the stellar mass density of the Universe based on the integrated \oii~SFRD. Overlaid are the SMD measurements from the literature that were compiled in the recent review of \citet{Madau2014}. We find that our integration of the purely \oii~ determined SFRD reasonably traces the stellar mass assembly of the Universe.}
\label{fig:smd}
\end{figure}

Our results are shown in figure \ref{fig:smd}. We find an evolution where the stellar mass assembly rapidly increases from $10^{6.2}$ to $10^8 ~\mathrm{M}_\odot ~ \mathrm{Mpc}^{-3}$ from $z \sim 5$ to $2$, a time frame of only 2 Gyr. The evolution then tapers and flattens out by $z = 0$ which is related to the decrease in the SFRD that we have observed since $z \sim 2$. This is also the same conclusion found by observational studies of the SMD. We include measurements from \citet{Arnouts2007}, \citet{Elsner2008}, \citet{Gallazzi2008}, \citet{Perez-Gonzalez2008}, \citet{Kajisawa2009}, \citet{Li2009}, \citet{Marchesini2009}, \citet{Yabe2009}, \citet{Pozzetti2010}, \citet{Caputi2011}, \citet{Gonzalez2011}, \citet{Bielby2012}, \citet{Lee2012}, \citet{Reddy2012}, \citet{Ilbert2013}, \citet{Moustakas2013}, and \citet{Muzzin2013} in figure \ref{fig:smd} and we find that our integrated SFRD reproduces the same evolution seen by these studies. We have found that the \oii~based SFRD and SMD accurately reproduce the evolution of mass assembly in the Universe. This match can also be seen as yet another verification that our sample of \oii~emitters are primarily star-forming galaxies as the conclusions from our SMD estimate are the same seen in the literature.

\section{Conclusions}
We have presented the largest sample of \hb~and \oii~emitters between $z \sim 0.8 - 5$ that have been selected based on a robust and self-consistent technique, backed up by a wide array of spectroscopic emitters. We have used the HiZELS UKIRT and Subaru narrow-band catalogs, along with multi-wavelength data from the COSMOS and UDS fields, to create a clean and well-defined sample of star-forming galaxies. The main results of this paper are as follows:\\
\indent (i) We have robustly selected a total of 2477, 371 , 270, 179 \hb~emitters at $z = 0.84$, 1.42, 2.23, and 3.24 and 3285, 137, 35, 18 \oii~emitters at $z = 1.47$, 2.25, 3.34, and 4.69 in the combined COSMOS and UDS fields. These are the largest samples of \hb~and \oii~emitters to have been detected in this redshift range.\\
\indent (ii) We have extended the luminosity function in the literature to higher-$z$, as well as refined the lower-$z$ measurements for both types of emitters. For the \hb~emitters, we find that the bright-end of our $z = 1.42$ and $z = 2.23$ LFs are in agreement with the grism spectroscopy-based luminosity functions of \citet{Colbert2013}; hence, this increases the reliability of our sample being dominantly [O\rom{3}]~emitters in the bright-end. We also find from our predictions of the [O\rom{3}] LFs that our sample is dominated by [O\rom{3}] emitters at the bright-end. The faint-end is dominated by H$\beta$ emitters. We also find that the normalization of the [O\rom{3}] LFs are the same such that the relative contribution of H$\beta$ emitters is the same between $z \sim 0.8 - 2.2$. 
\indent (iii) The evolution of $L_\star$ and $\phi_\star$ for \hb~is found to have a strong increasing/decreasing evolution, respectively, up to $z\sim3$. For our \oii~sample, we find that $L_\star$ increases strongly up to $z \sim5$ and $\phi_\star$ is strongly dropping up to the same redshift.\\
\indent (iv) We have discussed that our luminosity functions are reliable to be used in making predictions of the number of emitters to be detected by future wide-surveys, such as {\it Euclid} and WFIRST. Furthermore, our luminosity functions can also determine the number of low-$z$ interlopers in Ly$\alpha$ studies, such that the level of contamination by low-$z$ sources can be reduced in such studies.\\
\indent (v) The SFRD has been constrained using \oii~measurements up to $z \sim 5$ for the first time. We find that the peak of the cosmic SFRD is located around $z \sim 3$ and is in agreement with our large compilation of UV, IR, radio, and nebular emission studies. We find that for $z > 2$, our SFRD fit drops slightly faster in comparison to the UV dropout studies in this redshift regime. However, we find that the UV measurements are within the 1-$\sigma$ error bar range of our SFRD fit. Future space-based narrow-band surveys, such as {\it JWST} and {\it WISH}, will be able to extend the range of \oii~detection out to $z \sim 12$ so that we can compare and confirm or invalidate the UV dropout measurements.\\
\indent (vi) We also find that the \hb~SFRD measurements are nicely in line with our \oii~sample and other star-formation tracers. This then brings to question of whether the \hb~calibration is more ``reliable'' as a tracer of star-formation than previously thought. With our large sample of these emitters, we will have the ability to explore this issue in detail.\\
\indent (vii) By integrating the SFRD, we have made estimates of the stellar mass density evolution and find that it steeply rose up to $z\sim2$ and flattened out up to the present-day. This is also confirmed by the wealth of measurements in the literature. 

The results in the paper have implications in the evolution of galaxies and the star-formation activity occurring in said galaxies. Despite the robustness of our sample, there is still room for improvement. Our measurements have done well to constrain the bright-end, while keeping the faint-end fixed based on measurements from the literature. We will require deeper narrow-band and broad-band measurements in order to constrain the faint-end slope of the LF.  Spectroscopic follow-up will also be necessary to accurately measure the extent of AGN contamination in our sample. Although, our color-color selections have shown (see figure \ref{fig:BRiK}) that they are quite reliable due to the large set of spectroscopic measurements confirming this reliability. That being said, spectroscopic measurements of our sample will help in separating the H$\beta$ and [O\rom{3}] samples to measure separate luminosity functions. Lastly, future narrow-band surveys, such as the proposed {\it WISH} telescope, will be able to extend the redshift window of \hb~and \oii~studies up to $z \sim 12$, which can be used to confirm the UV dropout studies at higher-$z$. Despite all these improvements and potential future progresses, our sample has reliably (given all the limitations) and robustly traced the evolution of star-forming activity in the universe. 

\section*{Acknowledgments}
We thank the anonymous referee for their informative, detailed, and useful comments/questions. We also acknowledge Anahita Alavi for many useful and insightful discussion regarding the determination of the luminosity function. We also acknowledge Brian Siana for useful comments.

The data used in this paper is publicly available from \citet{Sobral2013}. We refer the reader to this paper for details in regards to the data reduction and selection methodology for the original catalogs.

This paper uses data from the VIMOS Public Extragalactic Redshift Survey (VIPERS). VIPERS has been performed using the ESO Very Large Telescope, under the ``Large Programme" 182.A-0886. The participating institutions and funding agencies are listed at http://vipers.inaf.it

DS acknowledges financial support from the Netherlands Organisation for Scientific research (NWO) through a Veni fellowship, from FCT through a FCT investigator Starting Grant and Start-up Grant (IF/01154/2012/CP0189/CT0010) and from FCT grant PEst-OE/FIS/UI2751/2014

IRS acknowledges support from STFC (ST/L00075X), the ERC Advanced Investigator programme DUSTYGAL 321334, and a Royal Society/Wolfson Merit Award.

PNB acknowledges support from STFC.

\bibliography{oii_lf}

\appendix

\newpage

\section{Selection Technique}
\label{sec:selection_technique}
Here, we present, in detail, the selection of emitters that made it in to our sample. We also present the exact color-color selections that were applied in our work in table \ref{table:cc_sel}. 

\begin{table*}
\centering
\caption{Definitions of the Color-Color Selection}
\begin{tabular}{c c c c c}
\hline
\hline
Filter & Color-Color & Emitter & Redshift & Selection Criteria\\
\hline
NB921 & $BRiK$ & \hb~ & 0.84 & $-0.3 < (B-R) < 0.08$ \& $(i - K) < 2.04(B - R)+0.81$\\ 
& & & & $0.08 < (B-R) < 1$ \& $1.6(B-R) < (i-K) < 2.04(B-R)+0.81$\\
& & &  &$1< (B-R) < 1.24$ \& $1.6(B-R) < (i-K) < 3.21$\\
& & &  & $1.24< (B-R)$ \& $2.01 < (i - K) < 3.21$\\
& & \oii~ & 1.47 & $(B-R) < -0.3$\\
& & & & $-0.3 < (B-R) < 1.17$ \& $(i-K) > 2.04(B-R) + 0.81$\\
& & & & $ (B-R) > 1.17$ \& $(i-K) > 3.21$\\
\hline
NBJ & $BzK$ & \hb~\& \oii & & $(B-z) < 0.4$\\
& & & & $0.4 < (B - z) < 2.41$ \& $(z-K) > (B - z) - 0.4$\\
& & & & $2.41 < (B - z)$ \& $(z - K) > 2.0$\\
& $izK$ & \hb & 1.42 & $(z - K) < 5 (i - z) - 0.4$\\
& & \oii & 2.25 & $(z - K) > 5 (i - z) - 0.4$\\
\hline
NBH & $BzK$  & \hb & 2.23 & $(B-z) < 0.4$\\
& & & & $0.4 < (B - z) < 2.41$ \& $(z-K) > (B - z) - 0.4$\\
& & & & $2.41 < (B - z)$ \& $(z - K) > 2.0$\\
& $izK$ & \hb & 2.23 & $(z - K) > 5 (i - z) - 0.4$\\
& $UVz$ & \oii & 3.34 & $(U-V) > 1.2$ \& $(U-V) > 0.5(V-z) +1.2$ \& $(V-z) < 1.6$\\
\hline
NBK & $UVz$ & \hb & 3.24 & $(U-V) > 1.2$ \& $(U-V) > 0.5(V-z) +1.2$ \& $(V-z) < 1.6$\\
 & $Viz$ & \oii & 4.69 & $(V-i) > 1.2$ \& $(V-i) > 0.89(i-z) +1.2$ \& $(i-z) < 1.3$\\
\hline
\hline
\end{tabular}
\label{table:cc_sel}
\end{table*}

%%% NB921 COLOR-COLOR PLOT
\begin{figure}
\centering
\includegraphics{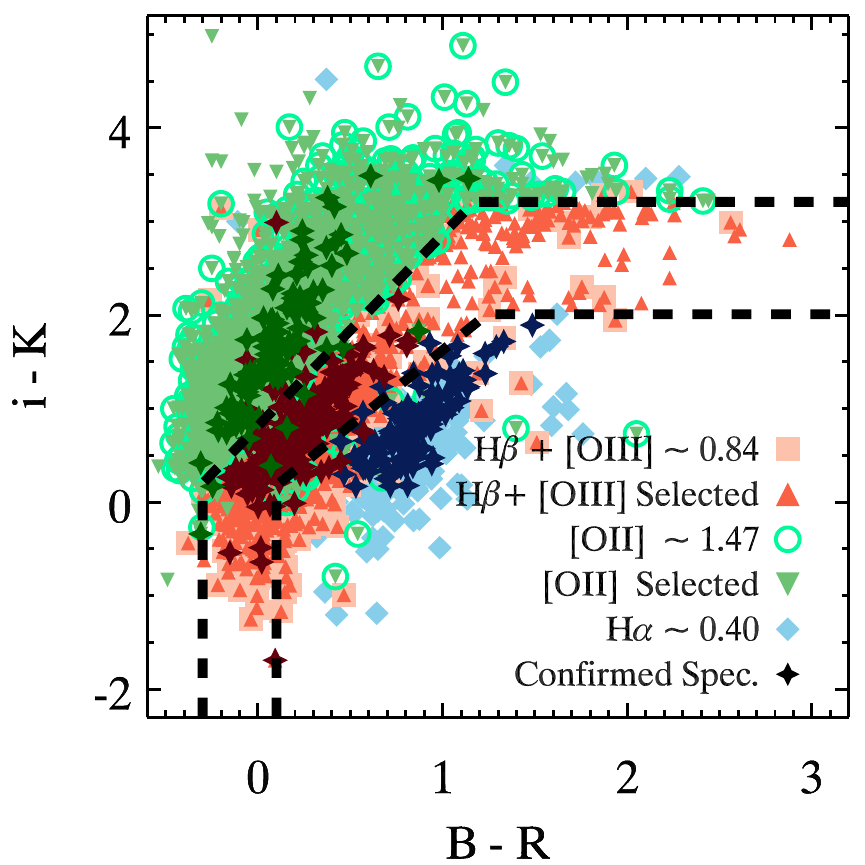}
\caption{The $i - K$ versus $B - R$ color-color distribution of NB921 emitters using the color-color selection of \citet{Sobral2013}. Overall, the level of completeness is $> 90\%$ as the large majority of spectroscopically confirmed emitters are within the selection area; hence, we confirm the the $BRiK$ selection of \citet{Sobral2009} is very efficient in selecting H$\alpha$, \hb, and \oii~samples from narrow-band surveys.
}
\label{fig:BRiK}
\end{figure}

\subsubsection{\hb~Emitters at $z \sim 0.8$}
\hb~sources in the NB921 data at $z \sim 0.8$ are selected by their photometric redshifts within the range of $0.75 < z_{phot} < 0.95$. The color-color selection criterion reduces the number of contaminants by separating the \hb~emitters from lower-$z$ H$\alpha$ and higher-$z$ \oii~emitters. This is done by using the $BRiK$ selection \citep{Sobral2009} in figure \ref{fig:BRiK}. Spectroscopic redshifts were used to assess the robustness and effectiveness of the selection criteria. 213 sources were spectroscopically confirmed. 169 were selected by the color-color selection. From the 213 sources (for which all were selected by their photo-$z$), only 11 were confirmed H$\beta$4861, 76 were [O\rom{3}]4959, and 126 were [O\rom{3}]5007 emitters. Removed from the sample where 9 low-$z$ and 13 high-$z$ spectroscopically confirmed emitters. The lower-$z$ emitters were primarily H$\alpha$ and [N\rom{2}]. The higher-$z$ emitters were primarily \oii~emitters. All these misidentified emitters were removed from the sample. Based on the spectroscopic data, we find that $\sim91\%$ of all spectroscopic measurements for photo-$z$ selected objects were either H$\beta$ or [O\rom{3}] emitters and that the $BRiK$ color-color selection does select $\sim 91\%$ of all the \hb~spectroscopically confirmed emitters. In total, we have 2477 $z = 0.84$ \hb~emitters in our sample. Based on the spectroscopic data, we find that the color-color selection effectively selects \hb~emitters with a completeness of $\sim 91\%$, making our sample not just the largest, but the most complete sample of \hb~emitters to date. 

\begin{figure*}
%\hspace{-2.5cm}
\includegraphics[scale=0.8]{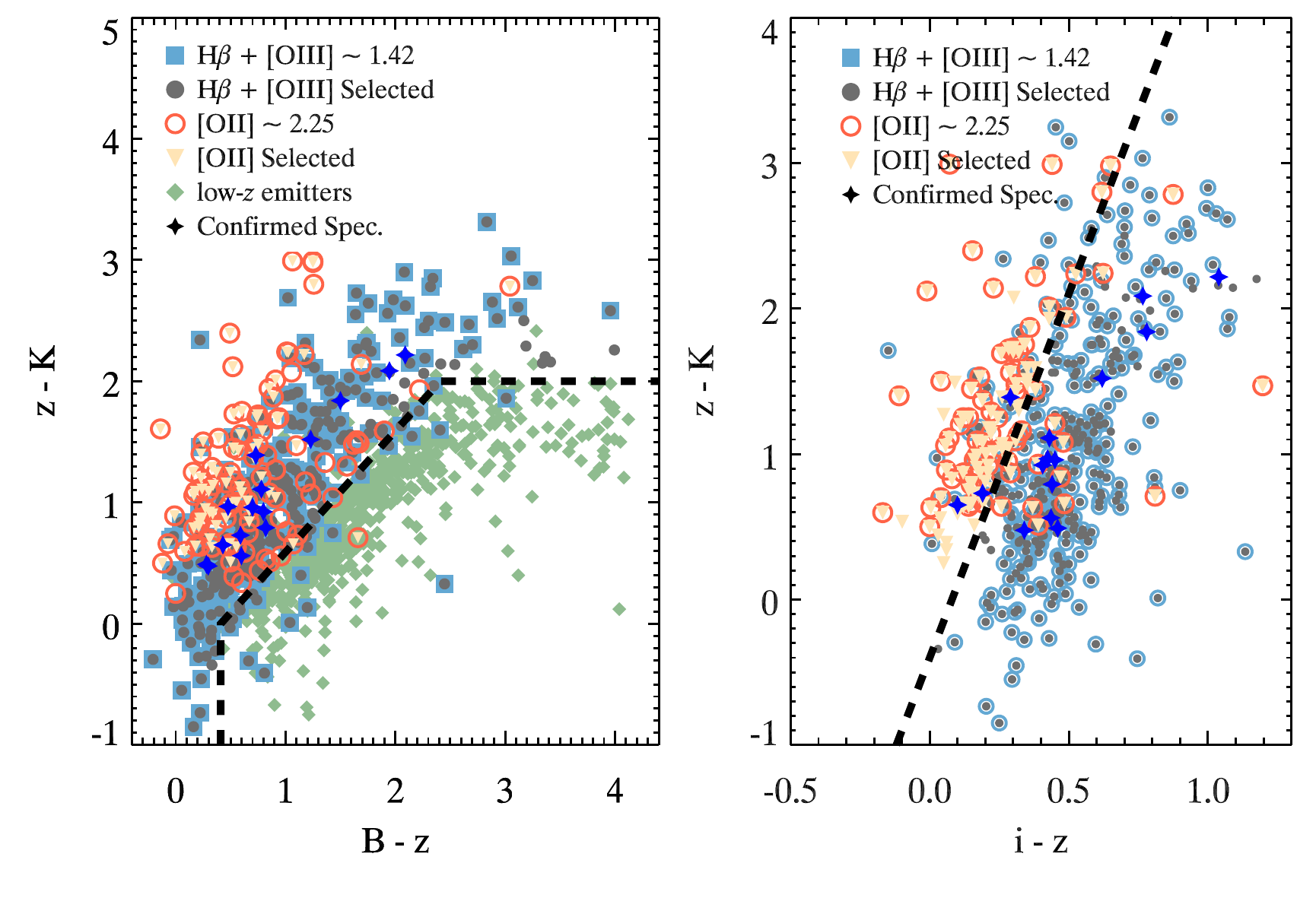}
\caption{Color-color magnitude distributions for all NBJ emitters. {\it Left:} The $z - K$ versus $B - z$ selection used to separate the \hb~ emitters at $z \sim 1.47$ and \oii~emitters at $z \sim 2.23$ from the low-$z$ emitters that are primarily H$\alpha$. {\it Right:} The $z - K$ versus $i - z$ selection is used to select \oii~ and \hb~emitters. For both selections, the spectroscopically confirmed emitters lie within the selection region adding to completeness of the sample. Both color-color selections nicely distinguish between the two samples. We note that for the $izK$ selection, about $\sim 15\%$ of selected \oii~emitters are within the selection region of \hb~emitters. These are photo-$z$ selected and shows that relying on purely the color-color selection would show a $\sim 15\%$ drop in the completeness of the \oii~sample, and a $\sim 15\%$ increase in the contamination of the \hb~sample.
}
\label{fig:NBJ}
\end{figure*}

\subsubsection{\hb~ \& \oii~Emitters at $z \sim 1.5$}
\oii~emitters in NB921 are selected with photometric redshifts between $1.2 < z_{phot} < 1.7$. We use the $BRiK$ color-color selection and include any sources with spectroscopic redshifts. Included in our sample are 97 spectroscopically confirmed sources, with 90 of them being color-color selected as well. Removed from the sample were 48 low-$z$ spectroscopically confirmed emitters, for which the majority were [O\rom{3}]. We also removed a few [He\rom{1}] emitters at $z\sim 1.28$ and H$\alpha$, and [N\rom{2}] emitters ($z\sim 0.4$). The contamination from high-$z$ emitters was significantly less (8 emitters) as there are no major emission lines beyond \oii. The majority of this contamination came from [Mg\rom{2}] emitters at $z = 2.25$. The issue of contamination arises here as we may ``naively'' state that our level of contamination is $\sim 33\%$, but it is noted that spectroscopic measurements to date have an inherent bias to the low-$z$ regime, such that there are more low-$z$ than high-$z$ measurements. This makes accurately measuring the level of contamination difficult. We note though that our color-color selection did select $\sim 93\%$ of the spectroscopically confirmed emitters. In total, we have selected 3285 $z = 1.47$ \oii~emitters. This is by far the largest sample of \oii~emitters at $z \sim 1.5$ to date and, based on the spectroscopically confirmed sources, is $\sim 93\%$ complete.

\hb~emitters in NBJ are selected based on a photometric redshift range of $1.20 < z_{phot} < 1.70$. The color-color selection criteria consists of a $BzK$ and $izK$ selection, as shown in figure \ref{fig:NBJ}. We use the $BzK$ selection to get our initial sample of emitters and remove the lower-$z$ contaminants, which are mostly H$\alpha$ emitters. To remove the higher-$z$ contaminants (\oii), we use the $izK$ selection. There were also 15 spectroscopically confirmed sources, all of which were within our color-color selection. Of these 15 emitters, 4 were H$\beta$, 5 [O\rom{3}]4958, and 6 [O\rom{3}]5007 emitters. We removed 5 emitters that were spectroscopically confirmed. These emitters were primarily H$\alpha$ and [N\rom{2}], all of which were removed from the sample. Our final sample consists of 371 \hb~ emitters at $z = 1.47$ that were selected.

\begin{figure*}
\includegraphics{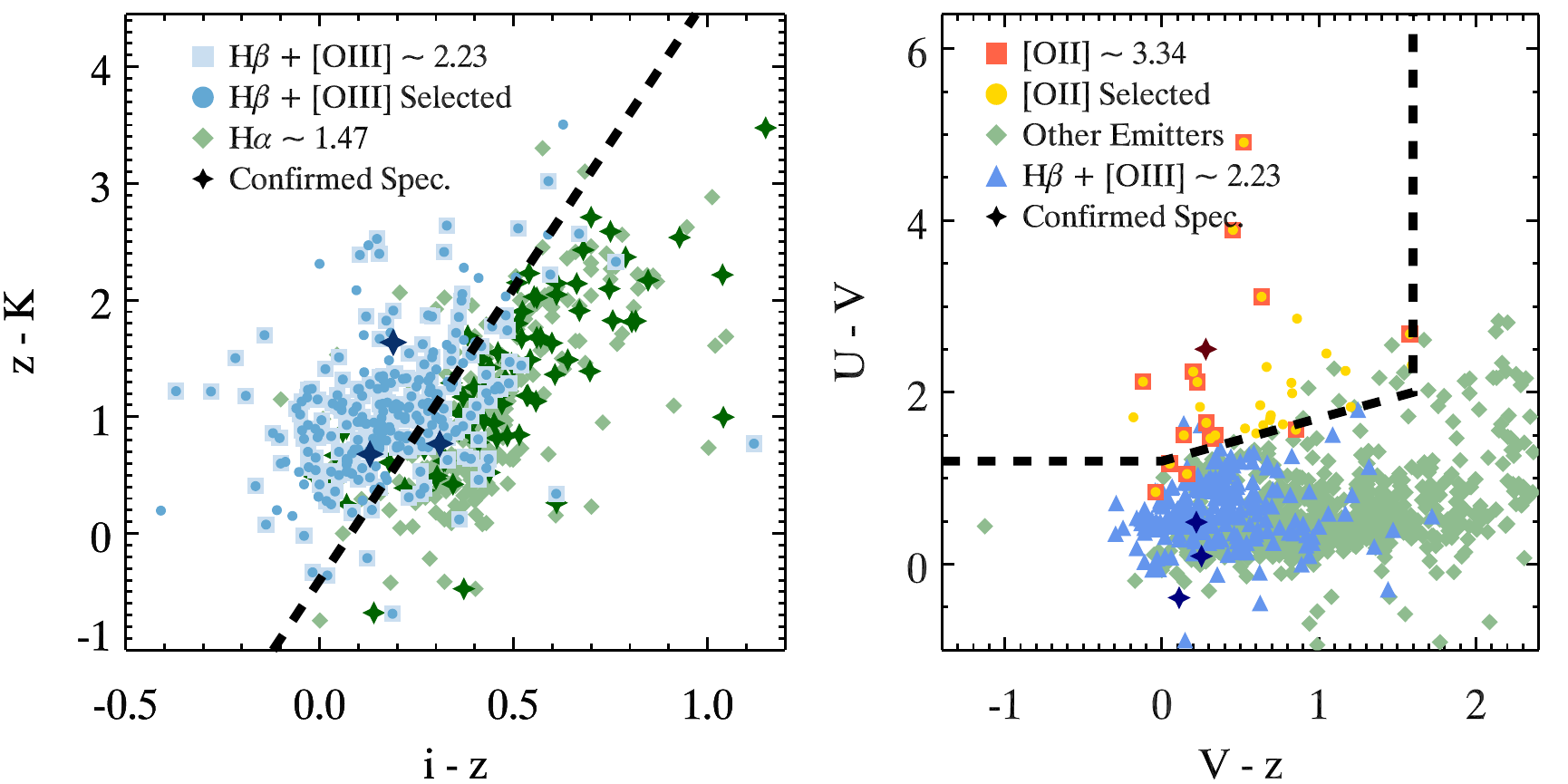}
\caption{Color-color magnitude distributions of all NBH emitters. {\it Left:} The $z - K$ versus $i - z$ color-color selection used to separate the \hb~$z\sim2.23$ emitters from the H$\alpha$ $z\sim1.47$ emitters (selected based on the methodology of \citealt{Sobral2013}). We have highlighted the spectroscopically confirmed emitters. The color-color selection shows a clear separation between H$\beta$ and H$\alpha$ measurements based on spectroscopic measurements alone. We find that $21\%$ of our photo-$z$ selected \hb~emitters are within the selection area for H$\alpha$ emitter, showing that relying on just color-color selection would result in the loss of $\sim 56$ emitters from the sample and an increase in contamination of the H$\alpha$ sample. {\it Right:} The $U - V$ versus $V-z$ color-color selection that is based on the Lyman break drop-out technique and is used to find $z\sim3.3$ \oii~emitters. We also include the \hb~sample and show that the vast majority of these emitters are outside the \oii~color-color selection region.}
\label{fig:NBH}
\end{figure*}

\begin{figure*}
\includegraphics{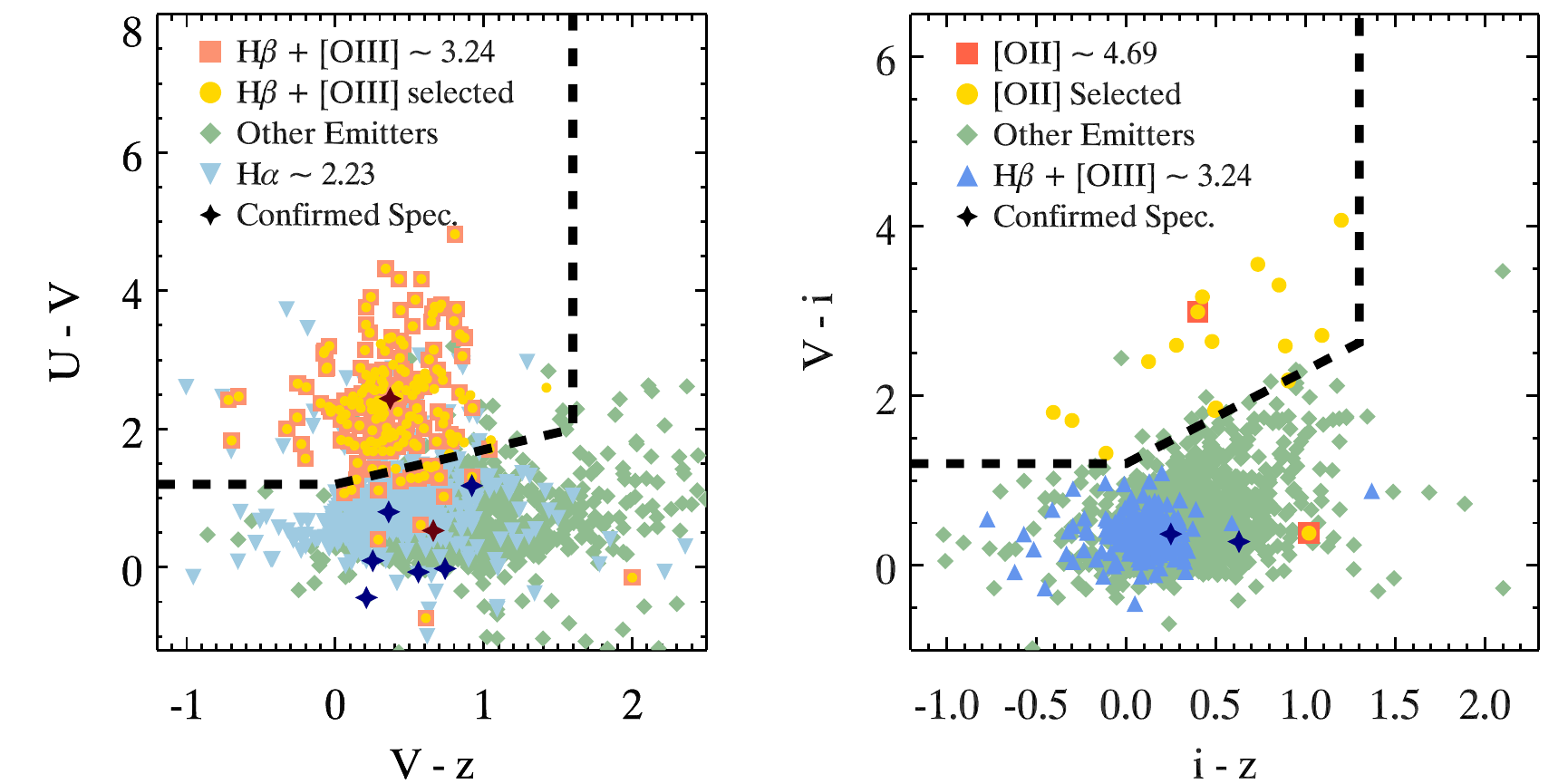}
\caption{Color-color magnitude distributions of all NBK emitters. {\it Left:} shows the $U-V$ versus $V-z$ color-color selection used to find $z\sim3.3$ \hb~emitters based on the Lyman break drop-out technique. We also include the H$\alpha$ emitters ($z\sim2.23$) and find that this selection reliably separates the two samples. {\it Right:} shows the $V-i$ versus $i-z$ color-color selection used to find $z\sim4.7$ \oii~emitters based also on the Lyman break drop-out technique. We include the \hb~sample to show that this selection criteria is reliable to separate the two samples. We find no \hb~emitter falling within the \oii~color-color region.}
\label{fig:NBK}
\end{figure*}

\subsubsection{\hb~ \& \oii~Emitters at $z \sim 2.2$}
\oii~emitters in NBJ are selected if their photometric redshifts are between $1.7 < z_{phot} < 2.8$. We apply the $BzK$ color-color selection to remove the lower-$z$ contaminants, which are primarily H$\alpha$ emitters, as shown in figure \ref{fig:NBJ}. We then use the $izK$ color-color selection to separate the sample from \hb~emitters. There were no spectroscopically confirmed sources included in the sample. Removed from the sample were 3 contaminants, which were all low-$z$ emitters (1 [N\rom{2}], 1 [O\rom{3}], and 1 [He\rom{1}]). In total, there are 137 \oii~emitters selected at $z = 2.25$.

\hb~emitters at $z\sim 2.2$ in NBH are selected with photometric redshifts between $1.7 < z_{phot} < 2.8$ along with $BzK$ and $izK$ color-color selections. We apply the $BzK$ color-color selection to remove emitters with $z \la 1.5$ (primarily the $z = 1.47$ H$\alpha$ emitters) and then apply the $izK$ color-color selection (shown in figure \ref{fig:NBH}) to separate our \hb~sample from the $z = 1.47$ H$\alpha$ emitters. We find that $21\%$ of our sample is outside the color-color region in figure \ref{fig:NBH} but is still selected via the photo-$z$ selection. This would raise concerns about the reliability of the color-color selection, but we must point out that it is reliable in terms of the separation between the \hb~and H$\alpha$ spectroscopically confirmed emitters. This also raises the point that we are more concerned with consistency in the use of selection techniques to reduce the effects of assumptions in our sample. Furthermore, this also shows that we can not rely on the color-color selection technique alone to select emitters since this would result in a 21\% drop in \hb~emitters and a 21\% increase in the contamination of the H$\alpha$ sample in \citet{Sobral2013}. By including the photo-$z$ in our selection methodology, we reduce the level of contaminants and also increase the reliability of the sample. We also included three spectroscopically confirmed sources in the sample that were selected by their color-color selection. Of these 3, we have two [O\rom{3}]4959 and one [O\rom{3}]5007 emitter. Removed from the sample were 3 low-$z$ spectroscopically confirmed emitters (H$\alpha$ at $z\sim 1.47$). There were no high-$z$ spectroscopic measurements that contaminated the sample. A total of 271 \hb~emitters at $z = 2.23$ were selected.

\subsubsection{\hb~ \& \oii~Emitters at $z \sim 3.3$}
\oii~emitters at $z = 3.3$ in NBH are selected if $2.8 < z_{phot} < 4$. We also select sources if they satisfy the $UVz$ color-color criteria. This separates the lower-$z$ contaminants from our sample. We include our \hb~sample in figure \ref{fig:NBH} to show the separation between the \oii~and \hb~sample and find that the $UVz$ selection is reliable in selecting our \oii~sample. Furthermore, as there are no major emission lines that are detected at higher-$z$, there is no need to include another color-color criteria to account for this contamination. As shown in fig. \ref{fig:redshift_distributions}, the number of sources greatly drops at higher-$z$ for the redshift of interest, making the number of contaminants very small. There was only one spectroscopically confirmed source that was included in the sample and only found in COSMOS. The $UVz$ color of this emitter places it well within the selection area, adding to the reliability of our color-color selection. No spectroscopic confirmed emitters were misidentified in the selection. In total, there are 35 of \oii~emitters at $z = 3.3$ selected.

Our \hb~emitters at $z=3.3$ in NBK are selected if their photometric redshifts lie between $2.8 < z_{phot} < 4$. We use a $UVz$ selection criteria, as shown in figure \ref{fig:NBH}, to select \hb~emitters based on the Lyman break dropout technique. Sources with no detection bluer than the $U$-band and detection in the $V$ and $z$ bands greater than the 5$\sigma$ magnitude detection limits were included. We also include all H$\alpha$ emitters with photo-$z$ around $z = 2.23$ in figure \ref{fig:NBK} to show the separation between them and our \hb~selected emitters. One spectroscopically confirmed source was also included in the sample from UDS. We find that this spectroscopically confirmed emitter is well within the color-color region. No spectroscopic confirmed emitters were misidentified in the selection. A total of 179 \hb~emitters at $z=3.3$ were selected.

\subsubsection{\oii~emitters at $z \sim 4.7$}
\label{sec:oii_NBK}
We select our \oii~emitters in NBK if they have a photometric redshift between $4.0 < z_{phot} < 6.0$. Emitters are also selected by using the $Viz$ criteria of \citet{Stark2009} for $V$-band dropouts. The color-color selection is shown in figure \ref{fig:NBK} and includes our \hb~sample. We find that there are no \hb~emitters within the \oii~selection area, which adds to the reliability of our sample. We find only one emitter was selected by its photo-$z$ within the selection range and it is well within the color-color selection region. The majority of emitters were selected by the color-color selection. All sources selected were also under the condition that anything bluer than $V$-band must have no detection. A study of \oii~emitters by \citet{Bayliss2012} also used a similar technique using the $BVz$ criteria of \citet{Stark2009} for their $z \sim 4.6$ sample. It must be noted though that the results of this study should not be taken as reliable due to the fact that they were limited to a sample size of only 3 \oii~emitters (about 3 times smaller than our sample) and the volume probed by this study is a factor of 100 times smaller than our study making it severely susceptible to cosmic variance. Our sample is statistically larger and robust in comparison to \citet{Bayliss2012}. There were no spectroscopically confirmed sources in our sample and no spectroscopic confirmed emitters were misidentified in the selection, thus, giving a total of 18 \oii~emitters at $z = 4.5$ that were selected.

\subsubsection{Notes on Contamination}
We advise the reader that the measurements of contamination are \emph{not} strictly reliable for the \hb~$z \sim 1.5$ emitters due to the bias in the spectroscopic redshift distribution. Two issues arise are: (1) lack of spectroscopic measurements at higher redshifts to properly quantify the level of contamination, and (2) the inherent bias of spectroscopic measurements to the lower-$z$ regime. The first point really just requires more spectroscopic measurements to increase the population of spectroscopically confirmed sources. The second point has to do with the distribution of spec-$z$ measurements. There exists more spec-$z$ measurements for $z < 1$, which results in a skewed histogram that favors the lower-$z$ regime. When measuring the level of contamination, there are more low-$z$ measurements than spectroscopically confirmed measurements (for example, $z = 1.47$ \oii) which causes a ``naive'' and biased estimation of the level of contamination. Instead, we considered where the spectroscopically confirmed \hb~and \oii~measurements were on the color-color diagrams as a way to assess the reliability of our selection technique.

Based on the points described above, we can measure the level of contamination for the \hb~$z \sim 0.84$ sample ($\sim 10\%$). To robustly measure the contamination for the higher-$z$ samples, we will need to conduct spectroscopic follow-up which is currently underway with Keck/MOSFIRE and ESO/VLT (Khostovan et al., in prep).

\newpage
\section{Binned Luminosity Function}
Here we include two tables that show the binned data points of the LF that are plotted in figures \ref{fig:hbeta} and \ref{fig:oii}. We include these plots as a convenience for future studies who wish to compare their LFs to ours.

\begin{table}
\caption{\hb~Luminosity Function. $\Phi_{\mathrm{obs}} $ shows the observed LF data points per bin; simply, it is the log of the number of emitters divided by the volume. $\Phi_{\mathrm{final}} $ is the completeness and filter profile corrected luminosity data points per bin. The errors here are Poissonian but with 20\% of the corrections added in quadrature.}
\resizebox{1.05\columnwidth}{!}{
\hspace{-1.0cm}
\begin{tabular}{c c c c c}
\hline
$\log_{10} L_{\mathrm{H}\beta + \mathrm{[O\rom{3}]}}$ & \#  & $\Phi_{\mathrm{obs}} $ & $\Phi_{\mathrm{final}}$ & Volume \\
(erg s$^{-1}$) & & (Mpc$^{-3} ~ \rm{d}\log_{10} \rm{L}$) & (Mpc$^{-3} ~ \rm{d}\log_{10} \rm{L}$) & ($10^5$ Mpc$^{3}$)\\
\hline
\hline
{\bf z = 0.84} & & & & \\
$41.10\pm0.10$&$703$&$-1.97$&$-1.82\pm0.02$&$3.25$\\
$41.30\pm0.10$&$465$&$-2.15$&$-2.04\pm0.03$&$3.25$\\
$41.50\pm0.10$&$262$&$-2.39$&$-2.35\pm0.04$&$3.25$\\
$41.70\pm0.10$&$128$&$-2.71$&$-2.61\pm0.06$&$3.25$\\
$41.90\pm0.10$&$68$&$-2.98$&$-2.94\pm0.08$&$3.25$\\
$42.10\pm0.10$&$28$&$-3.37$&$-3.17\pm0.13$&$3.25$\\
$42.30\pm0.10$&$12$&$-3.73$&$-3.52\pm0.20$&$3.25$\\
$42.50\pm0.10$&$3$&$-4.34$&$-4.12\pm0.39$&$3.25$\\
\hline
{\bf z = 1.42} & & & & \\
$41.95\pm0.15$&$284$&$-2.63$&$-2.49\pm0.03$&$4.06$\\
$42.25\pm0.15$&$73$&$-3.22$&$-3.14\pm0.07$&$4.06$\\
$42.55\pm0.15$&$12$&$-4.01$&$-3.89\pm0.19$&$4.06$\\
$42.85\pm0.15$&$2$&$-4.78$&$-4.64\pm0.48$&$4.06$\\
\hline
{\bf z = 2.23} & & & & \\
$42.60\pm0.075$&$84$&$-3.27$&$-3.08\pm0.06$&$10.46$\\
$42.75\pm0.075$&$70$&$-3.36$&$-3.14\pm0.07$&$10.69$\\
$42.90\pm0.075$&$22$&$-3.86$&$-3.65\pm0.13$&$10.69$\\
$43.05\pm0.075$&$5$&$-4.51$&$-4.26\pm0.29$&$10.69$\\
\hline
{\bf z = 3.24} & & & & \\
$42.65\pm0.075$&$70$&$-3.33$&$-3.17\pm0.07$&$9.99$\\
$42.80\pm0.075$&$52$&$-3.48$&$-3.26\pm0.09$&$10.48$\\
$42.95\pm0.075$&$25$&$-3.80$&$-3.55\pm0.13$&$10.48$\\
$43.10\pm0.075$&$6$&$-4.42$&$-4.17\pm0.27$&$10.48$\\
\hline
\hline
\end{tabular}
}
\label{table:hbeta_lf_observed}
\end{table}

\begin{table}
\caption{\oii~Luminosity Function. $\Phi_{\mathrm{obs}} $ shows the observed LF data points per bin; simply, it is the log of the number of emitters divided by the volume. $\Phi_{\mathrm{final}} $ is the completeness and filter profile corrected luminosity data points per bin. The errors here are Poissonian but with 20\% of the corrections added in quadrature.}
\resizebox{1.05\columnwidth}{!}{
\hspace{-1.0cm}
\begin{tabular}{c c c c c}
\hline
$\log_{10} L_{\mathrm{[O\rom{2}]}}$ & \#  & $\Phi_{\mathrm{obs}} $ & $\Phi_{\mathrm{final}}$ & Volume \\
(erg s$^{-1}$) & & (Mpc$^{-3} ~ \rm{d}\log_{10} \rm{L}$) & (Mpc$^{-3} ~ \rm{d}\log_{10} \rm{L}$) & ($10^5$ Mpc$^{3}$)\\
\hline
\hline
{\bf z = 1.47} & & & & \\
$41.65\pm0.075$&$590$&$-2.24$&$-2.08\pm0.02$&$6.80$\\
$41.80\pm0.075$&$425$&$-2.38$&$-2.28\pm0.03$&$6.80$\\
$41.95\pm0.075$&$257$&$-2.60$&$-2.46\pm0.04$&$6.80$\\
$42.10\pm0.075$&$127$&$-2.90$&$-2.69\pm0.06$&$6.80$\\
$42.25\pm0.075$&$42$&$-3.39$&$-3.05\pm0.10$&$6.80$\\
$42.40\pm0.075$&$19$&$-3.73$&$-3.55\pm0.15$&$6.80$\\
$42.55\pm0.075$&$6$&$-4.23$&$-4.23\pm0.28$&$6.80$\\
\hline
{\bf z = 2.25} & & & & \\
$42.45\pm0.10$&$92$&$-3.14$&$-2.77\pm0.05$&$6.29$\\
$42.65\pm0.10$&$37$&$-3.53$&$-3.15\pm0.08$&$6.29$\\
$42.85\pm0.10$&$3$&$-4.62$&$-4.46\pm0.35$&$6.29$\\
\hline
{\bf z = 3.34} & & & & \\
$43.05\pm0.050$&$12$&$-4.12$&$-3.86\pm0.17$&$15.88$\\
$43.15\pm0.075$&$7$&$-4.37$&$-3.92\pm0.24$&$16.52$\\
$43.30\pm0.075$&$2$&$-5.22$&$-4.87\pm0.48$&$16.52$\\
\hline
{\bf z = 4.69} & & & & \\
$42.86\pm0.075$&$10$&$-4.26$&$-3.66\pm0.09$&$12.22$\\
$43.01\pm0.075$&$5$&$-4.56$&$-3.93\pm0.13$&$12.22$\\
$43.16\pm0.075$&$2$&$-4.96$&$-4.11\pm0.16$&$12.22$\\
\hline
\hline
\end{tabular}
}
\label{table:oii_lf_observed}
\end{table}

\newpage
\section{Star-Formation Rate Density Compilation}
\label{sec:sfrd_comp}
In this section, we have compiled a table of the star-formation rate densities from different diagnostics spread over a wide redshift range. Because each study has its own set of assumptions, diagnostics, calibrations, dust corrections,etc. it is quite confusing in keeping track of which study has used which set of assumptions. Let alone, for the earliest papers, we have to even take into account the different cosmologies. To make life much easier for you as the reader who may be interested in studying the evolution of the cosmic SFR density, we have included in the appendix a long table which is our compilation of the SFR densities and luminosity function parameters from a range of different studies. Parts of this table are from \citet{Ly2007}, but updated with the newest studies in the field.

\begin{table*}
\caption{SFRD Compilation}
\begin{tabular}{l c c c c c c}
\hline
\hline
Study & z & Diagnostic & \multicolumn{3}{c}{Observed} & $\log_{10} \dot{\rho}_\star$\\
\cline{4- 6}
 & & & $\log_{10} \phi_\star$ & $\log_{10} L_\star$ & $\alpha$ & \\
 & & & Mpc$^{-3}$ & erg s$^{-1}$ & & M$_\odot$ yr$^{-1}$ Mpc$^{-3}$\\
 \hline
 %[OII] MEASUREMENTS
\citealt{Ciardullo2013} & $0.0 - 0.2$ & [O\rom{2}] & $-2.30^{+0.09}_{-0.11}$ & $40.32^{+0.18}_{-0.16}$ & $-1.2$ & $-2.05\pm{0.11}$\\
 & $0.2 - 0.325$ & [O\rom{2}] & $-2.12^{+0.05}_{-0.06}$ & $40.54\pm{0.11}$ & $-1.2$ & $-1.82\pm{0.06}$\\
 & $0.325 - 0.45$ & [O\rom{2}] & $-2.07^{+0.04}_{-0.05}$ & $40.75^{+0.08}_{-0.10}$ & $-1.2$ & $-1.71\pm{0.05}$\\
 & $0.45 - 0.56$ & [O\rom{2}] & $-2.07^{+0.03}_{-0.08}$ & $40.93^{+0.08}_{-0.12}$ & $-1.2$ & $-1.66\pm{0.06}$\\
\citealt{Sobral2012} & $1.47$ & [O\rom{2}] & $-2.01\pm0.10$ & $41.71\pm0.09$ & $-0.9\pm0.2$ & $-1.48\pm0.10$\\
\citealt{Bayliss2011} & 1.85 & [O\rom{2}] & $-2.23\pm0.09$ & $41.31\pm0.06$ & $-1.3\pm0.2$ & $-0.92\pm0.08$\\
\citealt{Ly2007} & $0.89$ & [O\rom{2}] & $-2.25\pm0.13$ & $41.33\pm0.09$ & $-1.27\pm0.14$ & $-1.68\pm{0.03}$\\
 & $0.91$ & [O\rom{2}] & $-1.97\pm0.09$ & $41.40\pm0.07$ & $-1.20\pm0.10$ & $-1.36\pm{0.02}$\\
 & $1.18$ & [O\rom{2}] & $-2.20\pm0.10$ & $41.74\pm0.07$ & $-1.15\pm0.11$ & $-1.27\pm{0.02}$\\
 & $1.47$ & [O\rom{2}] & $-1.97\pm0.06$ & $41.60\pm0.05$ & $-0.78\pm0.13$ & $-1.27\pm{0.02}$\\
\citealt{Zhu2009} & $0.84$ & [O\rom{2}] & ... & ... & ... & $-1.79^{+0.10}_{-0.10}$\\
 & $1.02$ & [O\rom{2}] & ... & ... & ... & $-1.75^{+0.13}_{-0.08}$\\
 & $1.19$ & [O\rom{2}] & ... & ... & ... & $-1.67^{+0.25}_{-0.11}$\\
 & $1.37$ & [O\rom{2}] & ... & ... & ... & $-1.60^{+0.16}_{-0.09}$\\
\citealt{Takahashi2007} & $1.71 - 1.203$ & [O\rom{2}] & $-2.37^{+0.10}_{-0.12}$ & $41.79^{+0.07}_{-0.06}$ & $-1.41^{+0.16}_{-0.15}$ & $-1.25^{+0.05}_{-0.08}$\\
   & $1.71 - 1.203$ & [O\rom{2}] & $-2.67^{+0.28}_{-0.49}$ & $41.75^{+0.32}_{-0.20}$ & $-1.38^{+0.40}_{-0.37}$ & $-1.61^{+0.09}_{-0.28}$\\
\citealt{Glazebrook2004} & $0.90$ & [O\rom{2}] & $-2.91$ & $42.30$ & $-1.3$ & $-1.35^{+0.34}_{-0.30}$\\
\citealt{Teplitz2003} & $0.90\pm0.50$ & [O\rom{2}] & $-3.06\pm0.12$ & $42.15\pm0.08$ & $-1.35$ & $-1.55\pm0.06$\\
\citealt{Gallego2002} & $0.025\pm0.025$ & [O\rom{2}] & $-3.48\pm0.19$ & $41.24\pm0.13$ & $-1.21\pm0.21$ & $-3.02\pm0.15$\\
\citealt{Hicks2002} & $1.20\pm0.40$ & [O\rom{2}] & ... & ... & ... & $-1.59^{+0.30}_{-0.48}$\\
\citealt{Hogg1998} & $0.20\pm0.10$ & [O\rom{2}] & ... & ... & ... & $-2.37^{+0.11}_{-0.16}$\\
  & $0.40\pm0.10$ & [O\rom{2}] & ... & ... & ... & $-1.77^{+0.09}_{-0.12}$\\
  & $0.60\pm0.10$ & [O\rom{2}] & ... & ... & ... & $-1.69^{+0.06}_{-0.08}$\\
  & $0.80\pm0.10$ & [O\rom{2}] & ... & ... & ... & $-1.75^{+0.07}_{-0.08}$\\
  & $1.00\pm0.10$ & [O\rom{2}] & ... & ... & ... & $-1.44^{+0.09}_{-0.11}$\\
  & $1.20\pm0.10$ & [O\rom{2}] & ... & ... & ... & $-1.57^{+0.18}_{-0.30}$\\ 
\citealt{Hammer1997} & $0.375\pm0.125$ & [O\rom{2}] & ... & ... & ... & $-2.20^{+0.07}_{-0.08}$\\
 & $0.625\pm0.125$ & [O\rom{2}] & ... & ... & ... & $-1.72^{+0.11}_{-0.15}$\\
 & $0.875\pm0.125$ & [O\rom{2}] & ... & ... & ... & $-1.35^{+0.20}_{-0.38}$\\
       
 %HBETA/[OIII] MEASUREMENTS
 \citealt{Colbert2013} & $0.7 - 1.5$ & [O\rom{3}] & $-3.19\pm0.09$ & $42.34\pm0.06$ & $-1.40\pm0.15$ & ...\\
 & $1.5 - 2.3$ & [O\rom{3}] & $-3.74\pm0.43$ & $42.91\pm0.37$ & $-1.67\pm0.78$ & ...\\
 & $0.7 - 1.5$ & [O\rom{3}] & $-3.28\pm0.09$ & $42.39\pm0.08$ & $-1.50$ & ...\\
 & $1.5 - 2.3$ & [O\rom{3}] & $-3.60\pm0.14$ & $42.83\pm0.11$ & $-1.50$ & ...\\
 \citealt{Pirzkal2013} & $0.5\pm0.4$ & [O\rom{3}] & $-2.58^{+0.09}_{-0.09}$ & $41.3^{+0.09}_{-0.09}$ & $-1.21^{+0.08}_{-0.07}$ & ... \\
 \citealt{Ly2007} & $0.41$ & [O\rom{3}] & $-2.55\pm0.25$ & $41.17\pm0.22$ & $-1.49\pm0.11$ & $-2.17\pm{0.06}$\\
 & $0.42$ & [O\rom{3}] & $-2.38\pm0.22$ & $41.11\pm0.24$ & $-1.25\pm0.13$ & $-2.31\pm{0.09}$\\
 & $0.62$ & [O\rom{3}] & $-2.58\pm0.17$ & $41.51\pm0.15$ & $-1.22\pm0.13$ & $-2.06\pm{0.05}$\\
 & $0.83$ & [O\rom{3}] & $-2.54\pm0.15$ & $41.53\pm0.11$ & $-1.44\pm0.09$ & $-1.73\pm{0.03}$\\

%HALPHA MEASUREMENTS
%\citealt{Colbert2013} & $0.3 - 0.9$ & H$\alpha$ & $-2.51\pm0.11$ & $41.72\pm0.09$ & $-1.27\pm0.12$ & ...\\
%& $0.9 - 1.5$ & H$\alpha$ & $-2.70\pm0.12$ & $42.18\pm0.10$ & $-1.43\pm0.17$ & ...\\
%\citealt{Ly2007} & $0.24$ & H$\alpha$ & $-2.98\pm0.40$ & $41.25\pm0.34$ & $-1.70\pm0.10$ & $-2.37\pm{0.08}$\\
% & $0.40$ & H$\alpha$ & $-2.40\pm0.14$ & $41.29\pm0.13$ & $-1.28\pm0.07$ & $-2.10\pm{0.05}$\\
%\citealt{Ly2011} & $0.81$ & H$\alpha$ & $-3.20\pm0.54$ & $43.00\pm0.52$ & $-1.6\pm0.19$ & $-0.96\pm0.18$\\
%Glazebrook et al. 1999 & $0.885\pm0.099$ & H$\alpha$ & ... & ... & ... & $-0.97\pm0.10$\\
%Yan et al. 1999 & $1.30\pm0.60$ & H$\alpha$ & $-2.89$ & $42.82$ & $-1.35$ & $-0.96^{+0.09}_{-0.11}$ \\
%Hopkins et al. 2000 & $1.25\pm0.55$ & H$\alpha$ & $-3.11\pm0.20$ & $42.87\pm0.11$ & $-1.60\pm0.12$ & $-1.00$\\
%Pascual et al. 2005 & $0.242\pm0.014$ & H$\alpha$ & ... & ... & ... & $-1.77^{+0.08}_{-0.09}$\\
%Tresse et al. 2002 & $0.73^{+0.37}_{-0.23}$ & H$\alpha$ & $-2.36\pm0.06$ & $41.98\pm0.06$ & $-1.31\pm0.11$ & $-1.37\pm0.05$\\
%Fujita et al. 2003 & $0.242\pm0.009$ & H$\alpha$ & $-2.62\pm0.34$ & $41.55\pm0.25$ & $-1.53\pm0.15$ & $-1.90^{+0.08}_{-0.17}$\\
%Glazebrook et al. 2004 & $0.384\pm0.006$ & H$\alpha$ & $-2.55$ & $41.49$ & $-1.30$ & $-2.05^{+0.14}_{-0.21}$\\
%& $0.458\pm0.006$ & H$\alpha$ & $-2.23$ & $42.15$ & $-1.30$ & $-1.07^{+0.17}_{-0.14}$\\

 \hline
 \hline
 \end{tabular}
 \end{table*}

\bsp

\label{lastpage}

\end{document}